\documentclass[aps,superscriptaddress,eqsecnum,nofootinbib,preprintnumbers]{revtex4}
\usepackage{graphicx,epsfig}
\usepackage{amsmath}
\usepackage {amssymb}
\usepackage{subfigure}

\usepackage{relsize}

\newcommand{\be}{\begin{equation}}
\newcommand{\ee}{\end{equation}}
\newcommand{\bea}{\begin{eqnarray}}
\newcommand{\eea}{\end{eqnarray}}
\newcommand{\nn}{\nonumber}

\begin{document}

\title{5-dimensional braneworld with gravitating Nambu-Goto matching
conditions}

\author{Georgios Kofinas}\email{gkofinas@aegean.gr}
\affiliation{Research Group of Geometry, Dynamical Systems and
Cosmology\\
Department of Information and Communication Systems Engineering\\
University of the Aegean, Karlovassi 83200, Samos, Greece}

\author{Vasilios Zarikas}\email{vzarikas@teilam.gr}
\affiliation{Department of Electrical Engineering, ATEI Lamias,
35100 Lamia, Greece}

\date{\today}

\begin{abstract}
We continue the investigation of a recent proposal on alternative matching conditions for self-gravitating
defects which generalize the standard matching conditions. The reasoning for this study is the need for
consistency of the various codimension defects and the
existence of a meaningful equation of motion at the probe limit, things that seem to lack from the standard
approach. These matching conditions arise by varying the brane-bulk action with respect to the brane embedding
fields (and not with respect to the bulk metric at the brane position) in a way that takes into account the
gravitational back-reaction of the brane to the bulk. They always possess a Nambu-Goto probe limit and
any codimension defect is seemingly consistent for any second order bulk gravity theory. Here, we consider in
detail the case of a codimension-1 brane in five-dimensional Einstein gravity, derive the generic alternative
junction conditions and find the $Z_{2}-$symmetric braneworld cosmology, as well as its bulk extension.
Compared to the standard braneworld cosmology, the new one has an extra integration constant which
accounts for the today matter and dark energy contents, therefore, there is more freedom for accommodating the
observed cosmic features. One branch of the solution possesses the asymptotic linearized LFRW regime.
We have constrained the parameters so that to have a recent passage from a long deceleration era to a small
today acceleration epoch and we have computed the age of the universe, consistent with current data,
and the time-varying dark energy equation of state. For a range of the parameters it is possible for the
presented cosmology to provide a large acceleration in the high energy regime.

\end{abstract}

\maketitle

\section{Introduction} \label{Introduction}

Distributional (thin) branes in an appropriate dimensional spacetime model the dynamics of various physical
systems, as it is the universe itself. A classical infinitely thin test brane with tension (probe) moving in
a given background spacetime is governed at lowest order by the Nambu-Goto action \cite{CGC}. Variation of this
action with respect to the brane embedding fields gives the Nambu-Goto equations of motion which are geometrically
described by the vanishing of the trace of the extrinsic curvature, and therefore, the worldsheet swept by the
brane is extremal (minimal). When the gravitational field of the defect is taken into account both the bulk metric
and the brane position become dynamical. Here, we consider
throughout that the bulk metric is regular (finite and continuous) at the brane position.
The standard method for obtaining the equations of motion of a back-reacting brane is to consider the
bulk field equations with all the localized energy-momentum tensor included and to isolate and integrate out
the distributional terms. A discontinuous extrinsic curvature or a conical singularity can source such delta
functions of suitable codimension. While for thin shells Israel matching conditions are well-established
\cite{Israel 1966}, when the support of a generic distributional stress-energy tensor is higher-codimensional,
it does not make sense to consider solutions of Einstein's equations \cite{Israel 1977}, \cite{Geroch}, \cite{Garfinkle},
\cite{cline} (a pure brane tension is a special situation which is consistent \cite{Vilenkin}, \cite{Frolov}).

In \cite{Ruth 2004}, geometric junction conditions for a codimension-2 conical defect in six-dimensional
Einstein-Gauss-Bonnet theory were derived with the hope that the above inconsistency is not due to the
defect construction, but due to the inability of Einstein gravity to describe complicated distributional
solutions. In \cite{CKP}
the consistency of the whole set of junction plus bulk field equations was explicitly shown for an axially
symmetric codimension-2 cosmological brane in six-dimensional EGB gravity, and it is likely that the
consistency will remain for non-axial symmetry. Analogously, e.g. a 5-brane in eight dimensions is
again of codimension-2 and EGB theory would suffice, but for a 4-brane in eight dimensions (codimension-3)
the third Lovelock density \cite{Lovelock} would need for consistency. However, e.g. a 2-brane in six dimensions
is of codimension-3 and it is probably inconsistent since the spirit of the proposal is to include higher
Lovelock densities to accommodate higher codimension defects and there is no higher than the second
Lovelock density in six dimensions. In brief, the generalization of the proposal is that in a $D$-dimensional
spacetime the maximal $[(D\!-\!1)/2]$ Lovelock density should be included (possibly along with lower Lovelock
densities) and the branes with codimensions $\delta=1,2,...,[(D\!-\!1)/2]$ should be consistent according to
the standard treatment; for yet higher codimensions the situation is not clear and probably inconsistent.
In four dimensions the absence of higher Lovelock densities does not allow the existence of
generic codimension-2 or 3 defects, but even if four dimensions are not the actual spacetime dimensionality,
at certain length and energy scales it has been tested that four-dimensional Einstein gravity represents effectively
the spacetime to high accuracy, so a consistent four-dimensional framework would at least be desirable.
Beyond the above (possible) shortcomings, there are extra difficulties with handling distributional sources
inside an equation. For example, for a codimension-2 brane there are two kinds of distributions involved,
$\delta(r)/r$ and $\delta(r)$, where $r$ is the radial coordinate from the brane. If both distributions are used
to derive two matching conditions \cite{Soda}, then an unnatural and undesirable inconsistency for certain
boundary conditions arises \cite{CKP}. Moreover, there is the problem of the regularization of the
distributional equation, since multiplying by $r$ only one distributional term remains, while multiplying by
$r^{2}$ all distributions vanish. If however one considers the corresponding variational problem of brane-bulk
action (variation with respect to the bulk metric at the brane position), the volume element of integration
$rdrd\theta$ vanishes the $\delta(r)$ distribution and only one matching condition arises, consistent with
bulk dynamics.

Having stated the question of consistency of ``high'' codimension defects in either $D$ or 4-dimensional spacetime,
we now pass to the question of the probe limit. We note that the standard equations of motion of a self-gravitating
defect do not obey the natural condition of continuous deformation from the probe limit equation of motion
(which is the Nambu-Goto equation). Indeed, the Israel matching conditions under vanishing of the brane energy-momentum
tensor give vanishing extrinsic curvature (geodesic motion), and similarly the probe limit of a codimension-1
brane in EGB gravity \cite{Germani} is another equation of motion, while the codimension-2 matching condition in EGB
theory \cite{Ruth 2004}, \cite{CKP}, \cite{Charmousis} is a third equation of motion. To set an analogy,
the linearized equation of motion of a point particle in four dimensions \cite{MiSaTa}, \cite{QuWa} (which is not
involved in our discussion since in this case the bulk metric diverges on the brane) is a correction of the
geodesic equation of motion on a given background (of course, for a 0-brane the geodesic equation coincides with
the Nambu-Goto) and for a two-body system the probe limit is realized when one mass is much smaller than the other.
However, in an analogous case it has been shown \cite{GerochJang}, \cite{Ehlers} that a probe point mass moves on
the geodesic of a background-solution of \textsl{any} gravitational theory, so the above variety of probe equations of
motion for different gravitational theories (or also the dependence of the equation of motion on the codimension of the
defect) maybe is not acceptable. Additionally, the matching conditions in EGB or Lovelock gravity \cite{Germani},
\cite{Ruth 2004}, \cite{Charmousis} are cubic or quadratic algebraic equations in the extrinsic curvature with
the total brane energy-momentum tensor on their right-hand side. Switching off this brane content the probe
limit arises which is a cubic or quadratic equation, therefore, in general, it possesses a multiplicity of probe
solutions. This means that these theories do not predict according to the standard approach a unique equation
of motion at the probe level \footnote{this argument was mentioned to us by J. Zanelli}. In this spirit, a correct
probe limit equation of motion should be linear in the extrinsic curvature and such are the geodesic or the Nambu-Goto
equations.

We would like to finish the discussion on the standard approach mentioning another possible deficiency.
The variation of a bulk action with respect to the bulk metric, beyond the main bulk terms gives as usual additional
``garbage'' $D-$dimensional terms. With the exception of codimension-1 case where the inclusion of the Gibbons-Hawking
term on the hypersurface cancels these terms, for all higher codimension defects such terms cannot cancel whatever
terms are added on the defect. The only possible thing one could imagine is to consider a ``tube'' around the defect
(two planes for codimension-1, a tube for codimension-2, a sphere for codimension-3, etc.), convert the unpleasant terms
into ``tube'' terms, make the cancelation on the ``tube'', and take the shrink limit. The variation of the brane-bulk
action in the interior of the ``tube'', considering also the relevant distributional terms and integrating out
around the defect, will give the brane equation of motion. Therefore, the metric variation outside the ``tube'' has
to be independent in order to get the bulk equations of motion, and also it has to be independent on the defect to get
the brane equation of motion. For the ``tube'' terms to cancel, some condition on the metric variation has to be
assumed on the ``tube'' (either Dirichlet-like if generalized Gibbons-Hawking terms \cite{Germani}, \cite{robin}
are included on the ``tube'', or Newmann-like). It seems that in the shrink limit these ``tube'' conditions will be
inconsistent with the independence of the brane metric variation.

A criticism against the standard approach in the lines of the above discussion was performed in \cite{kof-ira},
together with a proposal for obtaining alternative matching conditions called ``gravitating Nambu-Goto matching
conditions''. These arise by varying the brane-bulk action with respect to the brane position variables (embedding
fields). Although the brane energy-momentum tensor is still defined by the variation of the brane action with respect
to the induced metric, however, this tensor enters the new matching conditions in a different way
than before. Here, the distributional terms are still present, not inside a distributional differential equation
leading directly to inconsistencies at certain cases, but rather smoothed out inside an integration. In \cite{kof-ira},
it was shown in particular the consistency of the codimension-2 defect in EGB gravity according to these alternative
junction conditions, while the consistency of the codimension-2 limit of Einstein gravity \cite{KofTom} was also
discussed. Gravitating matching conditions aim to satisfy all the previous shortcomings of the standard conditions.
Four-dimensional Einstein gravity seems to be consistent for any codimension brane and the same seems also true either
for Einstein or any Lovelock extension for all higher spacetime dimensions $D$ (since the inclusion of the maximal
Lovelock density now is not crucial). These alternative matching conditions always have the Nambu-Goto probe limit,
independently of the gravitational theory considered, the dimensionality of spacetime or the codimensionality of the
defect. Finally, since the proposed equation of motion for the defect is decoupled from the bulk metric variation,
the ``outside'' problem (outside the ``tube'') is well-posed with a boundary Dirichlet or Newmann type of variation
on the ``tube'' (the central line is defined by an independent variation with respect to the embedding fields).
The proposed matching conditions generalize the standard matching conditions, and so, all the solutions of the
bulk equations of motion plus the conventional matching conditions are still solutions of the current system of equations.

In the present work we study the codimension-1 case in Einstein gravity. Our approach is reminiscent of the
``Dirac style'' variation performed in \cite{Davidson}, however our resulting matching conditions in the general
case do not coincide with the matching conditions derived in \cite{Davidson}. We have applied various methods in order to
confirm the derived result. Our main conceptual point, in view of the above arguments of consistency of the
various codimension defects and their probe limit, is that the gravitating Nambu-Goto matching conditions may
be close to the correct direction for deriving realistic matching conditions. We have applied these matching conditions
for a $Z_{2}$ cosmological brane without imposing any restriction about the bulk. The cosmology derived is
different than the cosmology derived in \cite{Davidson} (where the bulk was assumed to be AdS$_{5}$)
and the bulk space found here is not AdS$_{5}$. The set-up of the paper is as follows: In section
\ref{General arguments and introduction of the method} the method is introduced as an extension of the Nambu-Goto
variation for any codimension, so that the contribution from the gravitational back-reaction is included. In section
\ref{General setup, matching conditions and effective equations} the generic alternative junction conditions of a
codimension-1 brane in five-dimensional Einstein gravity are derived and manipulated together with the remaining
effective equations on the brane. Analogous equations hold for other codimension-1 branes in other spacetime dimensions,
but we choose the 3-brane as it can represent our world in the braneworld scenario. In section \ref{Cosmol} we
specialize to the cosmological configuration, integrate the brane system of equations and find the brane cosmology.
This cosmology has richer structure compared to the cosmology derived according to the standard conditions. In section
\ref{bulk sol} we find the bulk extension of the brane cosmology. In section \ref{inve} we investigate the
cosmological equations which provide interesting and realistic cosmological evolutions and study their
phenomenological implications. Finally, in section \ref{Conclusions} we conclude.

\section{A brief introduction of the method}
\label{General arguments and introduction of the method}

In order to get an idea how the proposed variation with respect to the embedding fields of the brane position
is performed we give in this section a brief account of the method for any codimension (for more details see
\cite{kof-ira}). However, the exact derivation of section
\ref{General setup, matching conditions and effective equations} in 5-dimensional spacetime is independent of
this section. We start with a general four-dimensional action
of the form
\begin{equation}
s_{4}\!=\!\int_{\Sigma}d^{4}\chi\sqrt{|h|}\,L(h_{ij})
\label{ashg}
\end{equation}
in a $D$-dimensional spacetime, where $L$ is any
scalar on $\Sigma$ built up from the induced metric $h_{ij}$. The brane coordinates are $\chi^{i}$
($i,j...$ are coordinate indices on the brane) and the bulk coordinates are $x^{\mu}$ ($\mu,\nu,...$ are
$D$-dimensional indices). In the present paragraph the bulk metric $\textsl{g}_{\mu\nu}$ is fixed and
non-dynamical, while the treatment of a back-reacted metric will be given in the next paragraph of this
section. The embedding fields are the external (bulk) coordinates of the brane, so they are some functions
$x^{\mu}(\chi^{i})$. Let the brane is deformed to another position described by the displacement vector
$\bar{\delta}x^{\mu}(x^{\nu})$ and the corresponding variation of the various quantities is denoted by
$\bar{\delta}_{x}$. The variation of the
tangent vectors on the brane $x^{\mu}_{\,\,,i}$ is $\bar{\delta}_{x}(x^{\mu}_{\,\,,i})
=(\bar{\delta}x^{\mu})_{,i}\equiv \bar{\delta} x^{\mu}_{\,\,,i}$. Since the bulk coordinates do not change,
the variation of $\textsl{g}_{\mu\nu}$ is
\begin{equation}
\bar{\delta}_{x}\textsl{g}_{\mu\nu}\!=\!\textsl{g}_{\mu\nu,\lambda}\bar{\delta} x^{\lambda}
\label{nksa}
\end{equation}
and the variation of $h_{ij}=\textsl{g}_{\mu\nu}\,x^{\mu}_{\,\,,i}\,x^{\nu}_{\,\,,j}$ is
\begin{equation}
\bar{\delta}_{x}h_{ij}=\textsl{g}_{\mu\nu,\lambda}x^{\mu}_{\,\,,i}x^{\nu}_{\,\,,j}\bar{\delta} x^{\lambda}
+\textsl{g}_{\mu\nu}x^{\mu}_{\,\,,i}\bar{\delta} x^{\nu}_{\,\,,j}+\textsl{g}_{\mu\nu}x^{\nu}_{\,\,,j}
\bar{\delta} x^{\mu}_{\,\,,i}=x^{\mu}_{\,\,,i}x^{\nu}_{\,\,,j}(\textsl{g}_{\mu\nu,\lambda}\bar{\delta}x^{\lambda}
+\textsl{g}_{\mu\lambda}\bar{\delta}x^{\lambda}_{\,\,,\nu}+\textsl{g}_{\nu\lambda}\bar{\delta}x^{\lambda}_{\,\,,\mu}).
\label{klsd}
\end{equation}
The variation of $s_{4}$ becomes
\begin{equation}
\bar{\delta}_{x}s_{4}\!=\!\int_{\Sigma}d^{4}\chi\sqrt{|h|}\,\tau^{ij}\bar{\delta}_{x}h_{ij},
\label{sjds}
\end{equation}
where
$\tau^{ij}\!=\!\frac{\delta L}{\delta h_{ij}}+\frac{L}{2}h^{ij}$. Substituting $\bar{\delta}_{x}h_{ij}$,
integrating by parts and imposing $\bar{\delta} x^{\mu}|_{\partial \Sigma}=0$, we get
\begin{eqnarray}
\bar{\delta}_{x}s_{4}&=&-2
\int_{\Sigma}d^{4}\chi\sqrt{|h|}\,\textsl{g}_{\mu\sigma}[(\tau^{ij}x^{\mu}_{\,\,,i})_{|j}+\tau^{ij}
\Gamma^{\mu}_{\,\,\,\nu\lambda}x^{\nu}_{\,\,,i}x^{\lambda}_{\,\,,j}]\bar{\delta} x^{\sigma}\nn\\
&=&-2
\int_{\Sigma}d^{4}\chi\sqrt{|h|}\,\textsl{g}_{\mu\sigma}(\tau^{ij}_{\,\,\,\,|j}\,x^{\mu}_{\,\,,i}-
\tau^{ij}
K^{\alpha}_{\,\,\,\,ij}n_{\alpha}^{\,\,\,\mu})\bar{\delta} x^{\sigma},
\label{dfjk}
\end{eqnarray}
since $x^{\mu}_{\,\,|ij}=x^{\mu}_{\,\,;ij}$,
$x^{\mu}_{\,\,;ij}+\Gamma^{\mu}_{\,\,\,\nu\lambda}x^{\nu}_{\,\,,i}x^{\lambda}_{\,\,,j}=-K^{\alpha}_{\,\,\,\,ij}
n_{\alpha}^{\,\,\,\mu}$, where $K^{\alpha}_{\,\,\,\,ij}=n^{\alpha}_{\,\,\,\,i;j}$ are the extrinsic curvatures on
the brane and $n_{\alpha}^{\,\,\,\mu}$ ($\alpha=1,...,\delta\!=\!D-4$) form a basis of normal vectors to the brane. The covariant
differentiations $|$ and $;$ correspond to $h_{ij}$ and $\textsl{g}_{\mu\nu}$ respectively, while
$\Gamma^{\mu}_{\,\,\,\nu\lambda}$ are the Christoffel symbols of
$\textsl{g}_{\mu\nu}$. Due to the arbitrariness of $\bar{\delta} x^{\mu}$ it arises
\begin{equation}
\tau^{ij}_{\,\,\,\,|j}\,x^{\mu}_{\,\,,i}-\tau^{ij}K^{\alpha}_{\,\,\,\,ij}n_{\alpha}^{\,\,\,\mu}=0
\label{snal}
\end{equation}
and since the vectors $x^{\mu}_{\,\,,i}$, $n_{\alpha}^{\,\,\,\mu}$ are independent, two sets of equations arise
\begin{equation}
\tau^{ij}_{\,\,\,\,|j}=0\,\,\,\,\,,\,\,\,\,\,
 \tau^{ij}K^{\alpha}_{\,\,\,\,ij}=0\Leftrightarrow\tau^{ij}(x^{\mu}_{\,\,;ij}
+\Gamma^{\mu}_{\,\,\,\nu\lambda}x^{\nu}_{\,\,,i}x^{\lambda}_{\,\,,j})=0.
\end{equation}
Note that the previous equivalence of the two expressions,
one with free index $\alpha$ and the other with free index $\mu$ is due to that the vectors
$x^{\mu}_{\,\,;ij}+\Gamma^{\mu}_{\,\,\,\nu\lambda}x^{\nu}_{\,\,,i}x^{\lambda}_{\,\,,j}$ are normal to the brane.
The variation described so far is the same with the one leading
to the Nambu-Goto equation of motion. Indeed for $L=1$, it is $\tau^{ij}=\frac{1}{2}h^{ij}$ and the
first equation is empty, while the second becomes $h^{ij}K^{\alpha}_{\,\,\,\,ij}=0
\Leftrightarrow \Box_{h}x^{\mu}+\Gamma^{\mu}_{\,\,\,\nu\lambda}h^{\nu\lambda}=0$
which is the Nambu-Goto equation of motion. Note again that the previous equivalence of the two expressions for the
Nambu-Goto equation, one with free index $\alpha$ and the other with free index $\mu$ is due to that the vector
$\Box_{h}x^{\mu}+\Gamma^{\mu}_{\,\,\,\nu\lambda}h^{\nu\lambda}$ is normal to the brane.
Similarly, the Regge-Teitelboim equation of motion \cite{Teitelboim} is a generalization where $L$ collects
the four-dimensional terms of (\ref{Stotal}), i.e. $L=\frac{r_{c}^{D-4}}{2\kappa_{D}^{2}}R-\lambda+\frac{L_{mat}}
{\sqrt{|h|}}$. It is $\tau^{ij}=-\frac{1}{2}\big(\frac{r_{c}^{D-4}}{\kappa_{D}^{2}}G^{ij}+\lambda h^{ij}
-T^{ij}\big)$, so the first equation becomes the standard conservation $T^{ij}_{\,\,\,\,\,|j}=0$ and the second
$\big(\frac{r_{c}^{D-4}}{\kappa_{D}^{2}}G^{ij}+\lambda h^{ij}-T^{ij}\big)K^{\alpha}_{\,\,\,\,ij}=0$.

In order to express the back-reaction of the brane onto the bulk and vice-versa, we
consider a general higher-dimensional action of the form
\begin{equation}
s_{_{\!D}}\!=\!\int_{M}d^{D}x\sqrt{|\textsl{g}|}\,\mathcal{L}(\textsl{g}_{\mu\nu}),
\label{skld}
\end{equation}
where $\mathcal{L}$
is any scalar on $M$ built up from the metric $\textsl{g}_{\mu\nu}$, e.g. $\mathcal{L}=\mathcal{R}(\textsl{g}
_{\mu\nu})$. Under an arbitrary variation of the bulk metric $\delta\textsl{g}_{\mu\nu}$ the variation of
$s_{_{\!D}}$ is
$\delta s_{_{\!D}}\!=\!\int_{M}d^{D}x\sqrt{|\textsl{g}|}\,\textsc{e}^{\mu\nu}\delta\textsl{g}_{\mu\nu}$,
where
$\textsc{e}^{\mu\nu}=\frac{\delta\mathcal{L}}{\delta\textsl{g}_{\mu\nu}}+\frac{\mathcal{L}}{2}\textsl{g}^{\mu\nu}$,
and the stationarity $\delta s_{_{\!D}}=0$ under arbitrary variations $\delta\textsl{g}_{\mu\nu}$ gives the bulk
field equations $\textsc{e}^{\mu\nu}=0$. The boundary terms arising from this variation in the presence of a
defect disappear by a suitable choice for the boundary condition of $\delta\textsl{g}_{\mu\nu}$,
usually by choosing a Dirichlet boundary condition for $\delta\textsl{g}_{\mu\nu}$.
However, in the presence of the defect, inside
$\textsc{e}^{\mu\nu}$, beyond the regular terms which obey $\textsc{e}^{\mu\nu}\!=\!0$, in general there are also
non-vanishing distributional terms making the variation $\delta s_{_{\!D}}$ not identically zero. The bulk
action knows about the defect through these distributional terms. Since
$\text{distr}\,\textsc{e}^{\mu\nu}\propto \delta^{(\delta)}$, where $\delta^{(\delta)}$ is the
$\delta$-dimensional delta function with support on the defect, it is
$(\text{distr}\,\textsc{e}^{\mu\nu})\delta\textsl{g}_{\mu\nu}\propto \delta^{(\delta)}\delta\textsl{g}_{\mu\nu}=
\delta^{(\delta)}\,\delta\textsl{g}_{\mu\nu}|_{brane}$, so only the variation of the bulk metric at the
brane position contributes to $\delta s_{_{\!D}}$, as expected.
More precisely, these distributional terms always appear in the parallel to the brane components and if
$\text{distr}\,\textsc{e}^{ij}=k^{ij}\delta^{(\delta)}$, the variation $\delta s_{_{\!D}}$ gets the form
\begin{equation}
\delta s_{_{\!D}}=\int_{M}d^{D}x\sqrt{|\textsl{g}|}\,k^{ij}\delta^{(\delta)}\delta h_{ij}=
\int_{\Sigma}d^{4}\chi\sqrt{|h|}\,k^{ij}\delta h_{ij}.
\label{djkj}
\end{equation}
Therefore, there is an extra variation of the bulk metric at the brane position $\delta\textsl{g}_{\mu\nu}|_{brane}$
(which in the adapted frame coincides with the variation of the induced metric $\delta h_{ij}$) which is
independent of the bulk
metric variation and this extra variation determines the brane equation of motion.
The corresponding variation of the total action at the brane position is
\begin{equation}
\delta (s_{_{\!D}}+s_{4})|_{brane}=\int_{\Sigma}d^{4}\chi\sqrt{|h|}\,(k^{ij}+\tau^{ij})\delta
h_{ij}.
\label{sdjk}
\end{equation}
In particular, if all the components of the variation $\delta h_{ij}$ are independent from each other,
the stationarity of the total action at the brane position $\delta (s_{_{\!D}}+s_{4})|_{brane}=0$ gives the
standard matching conditions (or standard brane equations of motion)
$k^{ij}+\tau^{ij}=0\Leftrightarrow k^{\mu\nu}\!+\!\tau^{\mu\nu}=0$, where $k^{\mu\nu}=k^{ij}x^{\mu}_{\,\,,i}
x^{\nu}_{\,\,,j}$, $\tau^{\mu\nu}=\tau^{ij}x^{\mu}_{\,\,,i}x^{\nu}_{\,\,,j}$ are parallel to the brane tensors.
Under a variation $\bar{\delta}x^{\mu}$ of the embedding fields, the variations of $\textsl{g}_{\mu\nu}$,
$h_{ij}$ were given in the previous paragraph and
the corresponding variation of the total action at the brane position will be
\begin{equation}
\bar{\delta}_{x}(s_{_{\!D}}+s_{4})|_{brane}=\int_{\Sigma}d^{4}\chi\sqrt{|h|}\,(k^{ij}+\tau^{ij})\bar{\delta}
_{x}h_{ij}.
\label{sjkl}
\end{equation}
Following the same steps as in the previous paragraph with $\tau^{ij}$ replaced by
$k^{ij}+\tau^{ij}$, the stationarity $\bar{\delta}_{x}(s_{_{\!D}}+s_{4})|_{brane}=0$ gives, due to the
arbitrariness of $\bar{\delta} x^{\mu}$, the brane equations of motion
\begin{equation}
(k^{ij}+\tau^{ij})_{|j}=0\,\,\,\,\,,\,\,\,\,\, (k^{ij}+\tau^{ij})K^{\alpha}_{\,\,\,\,ij}=0
\Leftrightarrow(k^{ij}+\tau^{ij})(x^{\mu}_{\,\,;ij}
+\Gamma^{\mu}_{\,\,\,\nu\lambda}x^{\nu}_{\,\,,i}x^{\lambda}_{\,\,,j})=0.
\label{klak}
\end{equation}
These can be called ``gravitating
Nambu-Goto matching conditions'' since they collect also the contribution from bulk gravity and they form a
schematic summary of our proposal. Nothing ab initio assures their consistency with the bulk field equations.
However, for the non-trivial case of a codimension-2 defect in six-dimensional EGB gravity the consistency has
been shown in \cite{kof-ira}. In order to describe
the previous variation of the brane position there is also the equivalent passive viewpoint of a bulk coordinate
change $x^{\mu}\rightarrow x'^{\mu}=x^{\mu}+\delta x^{\mu}$. Now the brane does not change position but it is
described by different coordinates and the change of the embedding fields is $\delta x^{\mu}$. Of course,
only the value of the variation $\delta x^{\mu}$ on the brane, and not the
values $\delta x^{\mu}$ away from the brane, is expected to influence the corresponding variation of the
brane-bulk action at the brane position. Although a bulk action $s_{_{\!D}}$ is invariant under coordinate
transformations, the presence of the defect, i.e. of the distributional terms inside $s_{_{\!D}}$, make
$\delta_{x}s_{_{\!D}}|_{brane}\!\neq\!0$. Let the tensor fields $\phi^{\mu}_{\nu}(x^{\rho})$
transform according to their {\textit{functional}} variation which is the change
in their functional form,
\begin{equation}
\delta_{x}\phi^{\mu}_{\nu}=\phi\,'^{\mu}_{\nu}(x^{\rho})-\phi^{\mu}_{\nu}(x^{\rho})
=\phi^{\lambda}_{\nu}(x^{\rho})\delta x^{\mu}_{,\lambda}
-\phi^{\mu}_{\lambda}(x^{\rho})\delta x^{\lambda}_{,\nu}-\phi^{\mu}_{\nu,\lambda}\delta x^{\lambda}
=-\pounds_{\delta x}\phi^{\mu}_{\nu},
\label{ksll}
\end{equation}
i.e. transform according to the Lie derivative with generator the
infinitesimal coordinate change. Therefore,
$\delta_{x} \textsl{g}_{\mu \nu}=
-(\textsl{g}_{\mu \nu , \lambda} \delta x^{\lambda}+\textsl{g}_{\mu
\lambda} {\delta x^{\lambda}}_{, \nu}+\textsl{g}_{\nu \lambda} {\delta
x^{\lambda}}_{,\mu})$,
while  $\delta_{x}x^{\mu}=0$, $\delta_{x}(x^{\mu}_{\,\,,i})=0$, thus
$\delta_{x}h_{ij}=(\delta_{x}\textsl{g}_{\mu\nu})x^{\mu}_{\,\,,i}x^{\nu}_{\,\,,j}$. The stationarity
under arbitrary variations $\delta x^{\mu}$ of the total brane-bulk action
$\delta_{x}(s_{_{\!D}}+s_{4})|_{brane}=\int_{\Sigma}d^{4}\chi\sqrt{|h|}\,(k^{ij}+\tau^{ij})\delta_{x}h_{ij}$
gives again the same matching conditions as before.

\section{Five-dimensional setup and alternative matching conditions}
\label{General setup, matching conditions and effective equations}

Let us consider the general system of five-dimensional Einstein gravity coupled to a localized 3-brane source.
The domain wall $\Sigma$ splits the spacetime $M$ into two parts $M_{\pm}$ and the two sides of $\Sigma$ are
denoted by $\Sigma_{\pm}$. The unit normal vector $n^{\mu}$ points inwards $M_{\pm}$. The total brane-bulk
action is
\begin{eqnarray}
S=\frac{1}{2\kappa_{5}^{2}}\!\int_{\!M}\!d^5x\sqrt{|\textsl{g}|}\,\big(\mathcal{R}-2\Lambda_{5}\big)
+\int_{\!\Sigma} \!d^4\chi\sqrt{|h|}\,\Big(\frac{r_{c}}{2\kappa_{5}^{2}}R-\lambda\Big)
-\frac{1}{\kappa_{5}^{2}}\!\int_{\!\Sigma_{\pm}} \!\!\!d^4\chi\sqrt{|h|}\,K
+\int_{\!M}\!d^5\!x\,\mathcal{L}_{mat}+\int_{\!\Sigma}\!d^{4}\!\chi\,L_{mat},\label{Stotal}
\end{eqnarray}
where $\textsl{g}_{\mu\nu}$ is the (continuous) bulk metric tensor and
$h_{\mu\nu}\!=\!\textsl{g}_{\mu\nu}-n_{\mu}n_{\nu}$ is the induced metric on the brane ($\mu,\nu,...$
are five-dimensional coordinate indices). The bulk coordinates are $x^{\mu}$ and the brane coordinates are
$\chi^{i}$ ($i,j,...$ are coordinate indices on
the brane). The symbol $\Sigma_{\pm}$ in an integral means the contribution from both sides of the surface. The
calligraphic quantities refer to the bulk metric, while the regular ones to the brane metric. The brane tension
is $\lambda$ (denoted also by $V$ in the next sections concerning cosmology) and the induced-gravity term
\cite{deffayet}, if present, has a crossover length scale
$r_{c}=2\kappa_{5}^{2}m^{2}=m^{2}/M^{3}$, where $\kappa_{5}^{-2}=M_{5}^{3}=2M^{3}=(8\pi G_{5})^{-1}$.
$\mathcal{L}_{mat}$, $L_{mat}$ are the matter Lagrangian densities of the bulk and of the brane respectively. The
contribution on each side of the wall of the Gibbons-Hawking term will also be necessary here as in the standard
treatment. $K=h^{\mu\nu}K_{\mu\nu}$ is the trace of the extrinsic curvature
$K_{\mu\nu}=h^{\kappa}_{\mu}h^{\lambda}_{\nu}n_{\kappa;\lambda}$ (the covariant differentiation $;$ corresponds
to $\textsl{g}_{\mu\nu}$).

Varying (\ref{Stotal}) with respect to the bulk metric we get the bulk equations of motion
\begin{equation}
\mathcal{G}_{\mu\nu}\!=\!\kappa_{5}^{2} \mathcal{T}_{\mu\nu}-\Lambda_{5}\textsl{g}_{\mu\nu}\,,
\label{bulk equations}
\end{equation}
where $\mathcal{G}_{\mu\nu}$ is the bulk Einstein tensor and $\mathcal{T}_{\mu\nu}$ is a regular bulk
energy-momentum tensor. We are mainly interested in a bulk with a pure cosmological constant
$\Lambda_{5}=\kappa_{5}^{2}\Lambda=\Lambda/2M^{3}$,
but for the present we leave a non-vanishing $\mathcal{T}_{\mu\nu}$. More precisely, we define the variation
$\delta\textsl{g}_{\mu\nu}$ of the bulk metric to vanish on the defect. In this variation, beyond the basic
terms proportional to $\delta\textsl{g}_{\mu\nu}$ which give (\ref{bulk equations}), there appear, as usually,
extra terms proportional to the second covariant derivatives $(\delta\textsl{g}_{\mu\nu})_{;\kappa\lambda}$
which lead to a surface integral on the brane with terms proportional to $(\delta \textsl{g}_{\mu\nu})_{;\kappa}$.
Adding the Gibbons-Hawking term, the normal derivatives of $\delta\textsl{g}_{\mu\nu}$, i.e. terms of the
form $n^{\kappa}(\delta\textsl{g}_{\mu\nu})_{;\kappa}$, are canceled. The remaining boundary terms are either
terms proportional to $\delta\textsl{g}_{\mu\nu}$ or terms containing
$h^{\mu\nu} n^{\lambda}(\delta\textsl{g}_{\mu\lambda})_{;\nu}$. These last terms lead (up to an irrelevant
integration on $\partial\Sigma$) again to terms proportional to $\delta\textsl{g}_{\mu\nu}$, and more precisely,
all the boundary terms together are basically of the known form
$(K^{\mu\nu}\!-\!Kh^{\mu\nu})\delta\textsl{g}_{\mu\nu}$.
Finally, considering as boundary condition for the variation of the bulk metric its vanishing on the brane
(Dirichlet boundary condition for $\delta\textsl{g}_{\mu\nu}$), there
is nothing left beyond the terms in equation (\ref{bulk equations}). The Gibbons-Hawking term will again
contribute in a while in another independent variation performed in order to obtain the brane equations of motion.
\par
According to the standard method, the interaction of the brane with the
bulk comes from the variation $\delta \textsl{g}_{\mu\nu}$ at the brane position of the action (\ref{Stotal}),
which is equivalent to adding on the right-hand side of equation (\ref{bulk equations}) the term $\kappa_{5}^2\,
\tilde{T}_{\mu\nu}\,\delta^{(1)}$, where
$\tilde{T}_{\mu\nu}\!=\!\sqrt{|h|/|\textsl{g}|}\,\big[T_{\mu\nu}\!-\!\lambda\,h_{\mu\nu}\!-\!(r_{c}/\kappa_{5}^2)G_{\mu\nu}
\big]$. $T_{\mu\nu}$ is the brane energy-momentum tensor, $G_{\mu\nu}$ the brane Einstein tensor and
$\delta^{(1)}$ the one-dimensional delta function with support on the defect. This approach leads to the
(generalized due to $r_{c}$) Israel matching conditions, it has been analyzed in numerous papers and discussed
in the Introduction.
\par
Here, we discuss an alternative approach where the interaction of the brane with bulk gravity is obtained
by varying the total action (\ref{Stotal}) with respect to $\delta x^{\mu}$, the embedding fields of the brane
position \cite{kof-ira}, \cite{Davidson}. The embedding fields are some functions $x^{\mu}(\chi^{i})$ and their
variations are $\delta x^{\mu}(x^{\nu})$. While in the standard method the variation of the bulk metric at
the brane position remains arbitrary, here the corresponding variation is induced by $\delta x^{\mu}$ as
explained in section \ref{General arguments and introduction of the method}, it is given by
\begin{equation}
\delta \textsl{g}_{\mu \nu}=\delta_{x}\textsl{g}_{\mu \nu}=\textsl{g}\,'_{\mu\nu}(x^{\rho})-\textsl{g}_{\mu\nu}
(x^{\rho})=-(\textsl{g}_{\mu\nu,\lambda}\delta x^{\lambda}+\textsl{g}_{\mu
\lambda}{\delta x^{\lambda}}_{,\nu}+\textsl{g}_{\nu\lambda}{\delta
x^{\lambda}}_{,\mu})=-\pounds_{\delta x} \textsl{g}_{\mu\nu}\,,
\label{delta x variation}
\end{equation}
and is obviously independent from the variation leading to (\ref{bulk equations}). The induced metric
$h_{ij}=\textsl{g}_{\mu\nu}\,x^{\mu}_{\,\,,i}\,x^{\nu}_{\,\,,j}$ enters the localized terms of the action
(\ref{Stotal}) and depends explicitly and implicitly (through $\textsl{g}_{\mu\nu}$) on the embedding fields.
Also the bulk terms of (\ref{Stotal}) contribute implicitly to the brane variation under the variation of the
embedding fields. The result of $\delta x^{\mu}$ variation gives, as we will see, as coefficient of $\delta x^{\mu}$
a combination of vectors parallel and normal to the brane, therefore, two sets of equations will finally arise
as matching conditions instead of one. Instead of directly expressing $\delta \textsl{g}_{\mu\nu}$,
$\delta h_{ij}$ in terms of $\delta x^{\mu}$, it is convenient to include the constraints in the action
and vary independently (however, we will also perform the direct calculation). So, the first constraint
$h_{ij}=\textsl{g}_{\mu\nu}\,x^{\mu}_{\,\,,i}\,x^{\nu}_{\,\,,j}$
implies the independent variation of $h_{ij}$. The variation $\delta x^{\mu}$ affects the variation of the
parallel to the brane vectors $x^{\mu}_{\,\,,i}$ which in turn influences the variation of the normal vector
$n^{\mu}$. So, the additional constraints $n_{\mu}x^{\mu}_{\,\,,i}=0$, $\textsl{g}_{\mu\nu}n^{\mu}n^{\nu}=1$
have to be added, and $\delta n_{\mu}$ is another independent variation. Finally, the third variation
$\delta \textsl{g}_{\mu\nu}$ depends on $\delta x^{\mu}$ by (\ref{delta x variation}). Therefore,
$\delta \textsl{g}_{\mu\nu}$ are independent from $\delta n_{\mu}$, $\delta h_{ij}$, but the various
$\delta \textsl{g}_{\mu\nu}$ components are not all independent from each other, so in the end they have to be
expressed in terms of $\delta x^{\mu}$ which are independent. If $\lambda^{ij}$, $\lambda^{i}$, $\lambda^{0}$ are
the Lagrange multipliers corresponding to the above constraints, the constraint action added to $S$ is
\begin{equation}
S_{c}=\int_{\Sigma_{\pm}}\!\!\!d^{4}\chi \,\sqrt{|h|} \,\Big [\lambda^{ij} (h_{ij}-\textsl{g}_{\mu\nu}
x^{\mu}_{\,\,,i} x^{\nu}_{\,\,,j})+\lambda^{i}\,n_{\mu} x^{\mu}_{\,\,,i}+\lambda^{0}\,(\textsl{g}_{\mu\nu}
n^{\mu} n^{\nu} -1)\Big]\,.
\label{Sconstraint}
\end{equation}
In general, the various Lagrange multipliers are different among the two sides $\Sigma_{\pm}$. Moreover, since
there are two normals $n^{\mu}_{+}, n^{\mu}_{-}$, there are two independent variations with respect to the
normals at the two sides. These two variations are independent since one could consider, for example, the case
where only the half space $M_{+}$ exists. Although the bulk metric $\textsl{g}_{\mu\nu}$ is continuous across
the defect, the variation $\delta\textsl{g}_{\mu\nu}$ is different among the two sides $\Sigma_{\pm}$. Indeed,
the extrinsic curvature is in general discontinuous on the brane and from equation (\ref{delta x variation}),
$\delta\textsl{g}_{\mu\nu}$ contains derivatives of the metric. Therefore, $\delta\textsl{g}_{\mu\nu}$ can
be expressed in terms of quantities on either side of the defect, but not simultaneously in terms of quantities
on both sides. Finally,
there is the extra independent variation $\delta h_{ij}$ with respect to the brane metric $h_{ij}$.

Variation of $S\!+\!S_{c}$ with respect to $n_{\mu},h_{ij},\textsl{g}_{\mu\nu}$ at the brane gives
\begin{eqnarray}
\delta(S\!+\!S_{c})\big|_{\!b\!r\!a\!n\!e}\!\!&=&\!\!
\int_{\!\Sigma_{\pm}}\!\!\!d^{4}\chi \sqrt{|h|}\,\Big(\!\lambda^{i}x^{\mu}_{\,\,,i}
\!+\!2\lambda^{0}n^{\mu}\!-\!\frac{1}{\kappa_{5}^{2}}Kn^{\mu}\!\Big)\delta n_{\mu}\nn\\
&+&\!\int_{\!\Sigma_{\pm}}\!\!\!d^{4}\chi \sqrt{|h|}\,\Big[\!\lambda^{ij}\!+\!\frac{1}{\kappa_{5}^{2}}
\Big(\!K^{ij}\!-\!\frac{K}{2}h^{ij}\!\Big)\!\Big]\delta h_{ij}
+\int_{\!\Sigma}\!\!d^{4}\chi \sqrt{|h|}\,\Big[\!\frac{1}{2}(T^{ij}\!-\!\lambda h^{ij})
\!-\!\frac{r_{c}}{2\kappa_{5}^{2}}G^{ij}\!\Big]\delta h_{ij}\nn\\
&-&\!\!\int_{\!\Sigma_{\pm}}\!\!\!d^{4}\chi\,\sqrt{|h|}\,\big(\lambda^{ij}x^{\mu}_{\,\,,i}x^{\nu}_{\,\,,j}
\!+\!\lambda^{0}n^{\mu}n^{\nu}\big) \delta\textsl{g}_{\mu\nu}\nn\\
&-&\!\!\frac{1}{2\kappa _{5}^{2}}\int_{\!M}\! d^{5}x \,\sqrt{|\textsl{g}|}
\,\Big \lbrace \mathcal{G}^{\mu\nu}\!-\!\kappa_{5}^{2}\mathcal{T}^{\mu\nu}\!+\!\Lambda_{5}\textsl{g}^{\mu\nu}
\!\Big \rbrace \delta \textsl{g}_{\mu \nu}\Big|_{\!b\!r\!a\!n\!e}
\!-\frac{1}{2\kappa_{5}^{2}}\int_{\!\Sigma_{\pm}}\!\!\!d^{4}\chi\,\sqrt{|h|}\, h^{\mu\nu} n^{\lambda}\,
\big[(\delta\textsl{g}_{\mu\lambda})_{;\nu}-(\delta\textsl{g}_{\mu\nu})_{;\lambda}\big]\nn\\
&+&\!\!\frac{1}{2\kappa_{5}^{2}}\int_{\!\Sigma_{\pm}}\!\!\!d^{4}\chi\, \sqrt{|h|}\, h^{\mu\nu} n^{\lambda}\,
\big[2(\delta\textsl{g}_{\mu\lambda})_{;\nu}-(\delta\textsl{g}_{\mu\nu})_{;\lambda}\big]\,.
\label{totalvariation}
\end{eqnarray}
When $r_{c}\!\neq\!0$, one should add in (\ref{Stotal}) the integral of the extrinsic curvature $k$ of
$\partial\Sigma$ (if $\partial\Sigma$ is not empty) to cancel some terms from the variation $\delta R$; this,
in general, does not affect the dynamics of $\Sigma$ \cite{guven}.

The basic root of difficulty for deriving (\ref{totalvariation}) is the Gibbons-Hawking term $K$. Its treatment,
due to the imposition of the constraints (\ref{Sconstraint}), is different from the conventional treatment of the
variation $\delta\textsl{g}_{\mu\nu}$ described after equation (\ref{bulk equations}). In the first line of
(\ref{totalvariation}), the $\delta n_{\mu}$ terms arise from the appropriate terms of (\ref{Sconstraint}) and the
$K\!=\!h^{ij}K_{ij}$ term of (\ref{Stotal}), where the identities $K_{ij}\!=\!n_{i;j}\!=\!-n_{\mu}(x^{\mu}_{\,\,;ij}
\!+\!\Gamma^{\mu}_{\,\,\,\nu\lambda}x^{\nu}_{\,\,,i}x^{\lambda}_{\,\,,j})$,
$-K_{ij}n^{\mu}= x^{\mu}_{\,\,;ij}+\Gamma^{\mu}_{\,\,\,\nu\lambda}x^{\nu}_{\,\,,i}x^{\lambda}_{\,\,,j}$ have been
used ($\Gamma^{\mu}_{\,\,\,\nu\lambda}$ are the Christoffel symbols of $\textsl{g}_{\mu\nu}$). These identities
show that $K_{ij}$ is a function of the independent variables $n_{\mu},\textsl{g}_{\mu\nu}$, but it does not
depend on the independent variable $h_{ij}$. Next, the $\delta h_{ij}$ terms in the second line of
(\ref{totalvariation}) arise again from the appropriate terms of (\ref{Sconstraint}) and all the four-dimensional
terms of (\ref{Stotal}). Finally, the variation $\delta\textsl{g}_{\mu\nu}$ on the brane is more difficult. The
first contribution comes from the appropriate terms of (\ref{Sconstraint}) and the derived terms are those of the
third line of (\ref{totalvariation}), where special care is needed for the variation of
$\textsl{g}_{\mu\nu}n^{\mu}n^{\nu}$ since $n_{\mu}$ is kept fixed and not $n^{\mu}$. Second, it arises from the
five-dimensional terms of (\ref{Stotal}) and the derived terms are those of the fourth line of
(\ref{totalvariation}). Third, it arises from the $K$ term of (\ref{Stotal}) and the derived terms are those of
the fifth line of (\ref{totalvariation}). In order to get these last terms, the identity
$K\!=\!h^{ij}n_{\mu,(i}\,x^{\mu}_{\,\,,j)}\!-\!n_{\mu}\Gamma^{\mu}_{\,\,\,\nu\lambda}h^{\nu\lambda}$ was used,
which arises from $K_{ij}\!=\!n_{\mu;\nu}x^{\mu}_{\,\,,i}x^{\nu}_{\,\,,j}\!=\!n_{\mu,(i}\,x^{\mu}_{\,\,,j)}
\!-\!n_{\mu}\Gamma^{\mu}_{\,\,\,\nu\lambda}x^{\nu}_{\,\,,i}x^{\lambda}_{\,\,,j}$ and
$h^{\mu\nu}=h^{ij}x^{\mu}_{\,\,,i}x^{\nu}_{\,\,,j}$.

The last term of the fourth line of (\ref{totalvariation}) cancels the last term of the fifth line, so the normal
derivatives of $\delta\textsl{g}_{\mu\nu}$ cancel. Then, use is made of the identity $h^{\mu\nu}n^{\lambda}
(\delta\textsl{g}_{\mu\lambda})_{;\nu}=(h^{\mu\nu}n^{\lambda}\delta\textsl{g}_{\mu\lambda})_{|\nu}-
(K^{\mu\nu}\!-\!Kn^{\mu}n^{\nu})\delta\textsl{g}_{\mu\nu}$, where $|$ denotes covariant differentiation with respect
to $h_{\mu\nu}$ (or $h_{ij}$), to convert the remaining $(\delta\textsl{g}_{\mu\lambda})_{;\nu}$ terms of
(\ref{totalvariation}) to $\delta\textsl{g}_{\mu\nu}$ terms (up to an irrelevant integration on $\partial \Sigma$).
Finally, the quantity in curly brackets appearing in the fourth line of (\ref{totalvariation}) vanishes since it
coincides with equation (\ref{bulk equations}) which is also valid on the brane.
The variation (\ref{totalvariation}) takes the form
\begin{eqnarray}
\delta(S\!+\!S_{c})\big|_{\!b\!r\!a\!n\!e}\!\!&=&\!\!\int_{\!\Sigma_{\pm}}\!\!\!d^{4}\chi \sqrt{|h|}\,
\Big(\!\lambda^{i}x^{\mu}_{\,\,,i}\!+\!2\lambda^{0}n^{\mu}\!-\!\frac{1}{\kappa_{5}^{2}}Kn^{\mu}\!\Big)
\delta n_{\mu}\nn\\
&+&\!\int_{\!\Sigma_{\pm}}\!\!\!d^{4}\chi \sqrt{|h|}\,\Big[\!\lambda^{ij}\!+\!\frac{1}{\kappa_{5}^{2}}
\Big(\!K^{ij}\!-\!\frac{K}{2}h^{ij}\!\Big)\!\Big]\delta h_{ij}
+\int_{\!\Sigma}\!\!d^{4}\chi \sqrt{|h|}\,\Big[\!\frac{1}{2}(T^{ij}\!-\!\lambda h^{ij})
\!-\!\frac{r_{c}}{2\kappa_{5}^{2}}G^{ij}\!\Big]\delta h_{ij}\nn\\
&-&\!\!\int_{\!\Sigma_{\pm}}\!\!\!d^{4}\chi\,\sqrt{|h|}\,\big(\lambda^{ij}x^{\mu}_{\,\,,i}x^{\nu}_{\,\,,j}
\!+\!\lambda^{0}n^{\mu}n^{\nu}\big) \delta
\textsl{g}_{\mu\nu}-\frac{1}{2\kappa _{5}^{2}}\int_{\!\Sigma_{\pm}}\!\!\!d^{4}\chi \,\sqrt{|h|}
\,\big(K^{\mu\nu}\!-\!Kn^{\mu}n^{\nu}\big)\delta\textsl{g}_{\mu\nu}\,.
\label{branevariation1}
\end{eqnarray}

As explained above, $\delta n_{\mu}, \delta h_{ij}$ are independent variations, but $\delta\textsl{g}_{\mu\nu}$
depends on $\delta x^{\mu}$ which are also independent. So, $\delta(S+S_{c})\big|_{\!b\!r\!a\!n\!e}=0$ gives
\begin{eqnarray}
&&\lambda^{i}
x^{\mu}_{\,\,,i}+\Big(2\lambda^{0}-\frac{1}{\kappa_{5}^{2}}K\Big)n^{\mu}=0
\label{n variation}\\
&&\lambda^{ij}_{+}+\lambda^{ij}_{-}+\frac{1}{\kappa_{5}^{2}}\Big(\!K^{ij}\!-\!\frac{K}{2}h^{ij}\!\Big)\Big|_{+}
+\frac{1}{\kappa_{5}^{2}}\Big(\!K^{ij}\!-\!\frac{K}{2}h^{ij}\!\Big)\Big|_{-}
+\frac{1}{2}\big(T^{ij}\!-\!\lambda h^{ij}\big)-\frac{r_{c}}{2\kappa_{5}^{2}}G^{ij}=0
\label{h variation}\\
&&\int_{\!\Sigma_{\pm}}\!\!\!d^{4}\chi\,\sqrt{|h|}\,\Big[\lambda^{ij}x^{\mu}_{\,\,,i}x^{\nu}_{\,\,,j}
+\lambda^{0}n^{\mu}n^{\nu}+\frac{1}{2\kappa _{5}^{2}}\big(K^{\mu\nu}\!-\!Kn^{\mu}n^{\nu}\big)\Big]
\delta\textsl{g}_{\mu\nu}=0\,,
\label{g variation}
\end{eqnarray}
where $\delta\textsl{g}_{\mu\nu}$ obeys (\ref{delta x variation}). Equation (\ref{n variation}) holds separately
for each side $\Sigma_{\pm}$. Since the vectors $x^{\mu}_{\,\,,i}$\,,\,$n_{\alpha}^{\,\,\,\mu}$ are independent,
equation (\ref{n variation}) implies for any side separately $\lambda^{i}=0$,
$\lambda^{0}=\frac{1}{2\kappa_{5}^{2}}K$. Equation (\ref{h variation}) contains the combination
$\lambda^{ij}_{+}+\lambda^{ij}_{-}$ and it will be seen that the matching conditions contain the same combination,
so the matching conditions will be unambiguously determined. Then, equation (\ref{g variation}), with
$\lambda^{ij}$ satisfying (\ref{h variation}), takes the form
\begin{equation}
\int_{\!\Sigma_{\pm}}\!\!\!d^{4}\chi\,\sqrt{|h|}\,\Big(\frac{1}{2\kappa _{5}^{2}}K^{\mu\nu}+\lambda^{ij}x^{\mu}_{\,\,,i}
x^{\nu}_{\,\,,j}\Big)\delta\textsl{g}_{\mu\nu}=0\,.
\label{diki}
\end{equation}
Since $K^{\mu\nu}=K^{ij}x^{\mu}_{\,\,,i}x^{\nu}_{\,\,,j}$, equation (\ref{diki}) is written as
\begin{equation}
\quad\quad\quad\quad\quad\quad\quad\quad
\int_{\!\Sigma_{\pm}}\!\!\!d^{4}\chi\,\sqrt{|h|}\,\mu^{ij}x^{\mu}_{\,\,,i}x^{\nu}_{\,\,,j}\,
\delta\textsl{g}_{\mu\nu}=0\quad\,\,\,,\,\,\,\quad
\mu^{ij}=\frac{1}{2\kappa _{5}^{2}}K^{ij}+\lambda^{ij}\,.
\label{fortsa}
\end{equation}
Contrary to the present situation, had we considered all $\delta n_{\mu}, \delta h_{ij}, \delta\textsl{g}_{\mu\nu}$
independent, equations (\ref{n variation}), (\ref{h variation}) would still arise. Equation
(\ref{fortsa}) would be written as
$\int_{\Sigma}d^{4}\chi\sqrt{|h|}\,(\mu^{ij}_{+}+\mu^{ij}_{-})x^{\mu}_{\,\,,i}x^{\nu}_{\,\,,j}\,
\delta\textsl{g}_{\mu\nu}=0$ providing $\mu^{ij}_{+}+\mu^{ij}_{-}=0$. Then, using equation (\ref{h variation}),
the Israel matching condition
$(K_{ij}-Kh_{ij})_{_{+}}+(K_{ij}-Kh_{ij})_{_{-}}=\kappa_{5}^{2}(\lambda h_{ij}-T_{ij})+r_{c}G_{ij}$
would arise.
\newline
In our approach $\delta\textsl{g}_{\mu\nu}$ has to be expressed via (\ref{delta x variation}) in terms of
quantities on either side and equation (\ref{fortsa}) becomes
\begin{equation}
\quad\quad\quad\quad\quad\quad\quad\quad
\int_{\!\Sigma_{+}\,\text{or}\,\Sigma_{-}}\!\!\!d^{4}\chi\,\sqrt{|h|}\,\mathcal{M}^{ij}x^{\mu}_{\,\,,i}x^{\nu}_{\,\,,j}\,
\delta\textsl{g}_{\mu\nu}=0\quad\,\,\,,\,\,\,\quad
\mathcal{M}^{ij}=\mu^{ij}_{+}+\mu^{ij}_{-}\,,
\label{pertouli}
\end{equation}
or equivalently
\begin{equation}
\int_{\!\Sigma_{+}\,\text{or}\,\Sigma_{-}}\!\!\!d^{4}\chi\,\sqrt{|h|}\,\mathcal{M}^{ij}
\big(\textsl{g}_{\mu\nu,\lambda}x^{\mu}_{\,\,,i}x^{\nu}_{\,\,,j}\delta
x^{\lambda}+2\textsl{g}_{\mu\nu}x^{\mu}_{\,\,,i}x^{\lambda}_{\,\,,j}\delta
x^{\nu}_{\,\,,\lambda}\big)\!=\!0\,. \label{trial}
\end{equation}
After an integration of (\ref{trial}) by parts and imposing $\delta x^{\mu}|_{\partial
\Sigma}=0$, we get
\begin{equation}
\int_{\!\Sigma_{+}\,\text{or}\,\Sigma_{-}}\!\!\!d^{4}\chi\,\sqrt{|h|}\,\textsl{g}_{\mu\sigma}
\big[\mathcal{M}^{ij}_{\,\,\,|j}\,x^{\mu}_{\,\,,i}
+\mathcal{M}^{ij}\big(x^{\mu}_{\,\,;ij}\!+\!\Gamma^{\mu}_{\,\,\,\nu\lambda}x^{\nu}_{\,\,,i}x^{\lambda}_{\,\,,j}\big)\big]
\delta x^{\sigma}=0
\label{step}
\end{equation}
and since the extrinsic curvature satisfies
$-K_{ij}n^{\mu}= x^{\mu}_{\,\,;ij}+\Gamma^{\mu}_{\,\,\,\nu\lambda}x^{\nu}_{\,\,,i}x^{\lambda}_{\,\,,j}$\,,
equation (\ref{step}) becomes
\begin{equation}
\int_{\!\Sigma_{+}\,\text{or}\,\Sigma_{-}}
\!\!\!d^{4}\chi\,\sqrt{|h|}\,\textsl{g}_{\mu\sigma}\big(\mathcal{M}^{ij}_{\,\,\,|j}\,
x^{\mu}_{\,\,,i}-\mathcal{M}^{ij}K_{ij}n^{\mu}\big)\delta x^{\sigma}=0\,.
\label{stepa}
\end{equation}
Due to the arbitrariness of $\delta x^{\mu}$ it holds
\begin{equation}
\mathcal{M}^{ij}_{\,\,\,|j}\,x^{\mu}_{\,\,,i}-\mathcal{M}^{ij}K_{ij}^{+}n^{\mu}_{+}=0\,\,\,,\,\,\,
\mathcal{M}^{ij}_{\,\,\,|j}\,x^{\mu}_{\,\,,i}-\mathcal{M}^{ij}K_{ij}^{-}n^{\mu}_{-}=0\,,
\label{naked}
\end{equation}
therefore, two sorts of matching conditions arise
\begin{eqnarray}
\mathcal{M}^{ij}K_{ij}^{+}=0\,&,&\,\mathcal{M}^{ij}K_{ij}^{-}=0\label{muu matching}\\
\mathcal{M}^{ij}_{\,\,\,|j}\!\!&=&\!\!0\,.
\label{mu matching}
\end{eqnarray}
Substituting $\lambda^{ij}_{+}+\lambda^{ij}_{-}$ of $\mathcal{M}^{ij}$ from (\ref{h variation}), we get
{\textit{the matching conditions of codimension-1 Einstein gravity}}
\begin{eqnarray}
&&\big[(K^{ij}_{+}-K_{+}h^{ij})+(K^{ij}_{-}-K_{-}h^{ij})+\kappa_{5}^{2}(T^{ij}-\lambda h^{ij})-r_{c}G^{ij}\big]
K_{ij}^{+}=0\label{matching1+}\\
&&\big[(K^{ij}_{+}-K_{+}h^{ij})+(K^{ij}_{-}-K_{-}h^{ij})+\kappa_{5}^{2}(T^{ij}-\lambda h^{ij})-r_{c}G^{ij}\big]
K_{ij}^{-}=0\label{matching1-}\\
&&\big[(K^{ij}_{+}-K_{+}h^{ij})+(K^{ij}_{-}-K_{-}h^{ij})+\kappa_{5}^{2}T^{ij}\big]_{|j}=0\,.
\label{matching2}
\end{eqnarray}
It is obvious that any solution of the standard matching conditions is still solution of the above matching
conditions, therefore, the new space of solutions is expected to be a continuous deformation of the space of
solutions of the standard theory.

If the extrinsic curvatures $K_{ij}^{+}$, $K_{ij}^{-}$ are proportional to each other, i.e.
$K_{ij}^{-}=\eta K_{ij}^{+}$, the previous matching conditions become
\begin{eqnarray}
&&\big[(1+\eta)(K^{ij}-K h^{ij})+\kappa_{5}^{2}(T^{ij}-\lambda h^{ij})-r_{c}G^{ij}\big]
K_{ij}=0\label{match1}\\
&&\big[(1+\eta)(K^{ij}-K h^{ij})+\kappa_{5}^{2}T^{ij}\big]_{|j}=0\,,
\label{match2}
\end{eqnarray}
where $K_{ij}\equiv K_{ij}^{+}$.
\newline
A $Z_{2}-$symmetric brane obeys $K_{ij}^{-}=K_{ij}^{+}\equiv K_{ij}$, so it corresponds to $\eta=1$. The
matching conditions become
\begin{eqnarray}
&&\Big[K^{ij}-Kh^{ij}+\frac{\kappa_{5}^{2}}{2}(T^{ij}-\lambda h^{ij})-\frac{r_{c}}{2}G^{ij}\Big]
K_{ij}=0\label{match1Z}\\
&&T^{ij}_{\,\,\,\,|j}=-\frac{2}{\kappa_{5}^{2}}\big(K^{ij}-Kh^{ij}\big)_{|j}\,.
\label{match2Z}
\end{eqnarray}
\newline
The ``smooth'' brane has $K_{ij}$ continuous, so $-K_{ij}^{-}=K_{ij}^{+}\equiv K_{ij}$ and
corresponds to $\eta=-1$. The matching conditions are
\begin{eqnarray}
&&\big[\kappa_{5}^{2}(T^{ij}-\lambda h^{ij})-r_{c}G^{ij}\big]K_{ij}=0\label{match1RT}\\
&&T^{ij}_{\,\,\,\,|j}=0\,.
\label{match2RT}
\end{eqnarray}

In \cite{Davidson}, the matching conditions derived are not the same with (\ref{matching1+})-(\ref{matching2}).
The reason is that although equations (\ref{n variation}), (\ref{h variation}) were still valid there, equation
$\int_{\Sigma_{+}}\!\!d^{4}\chi\sqrt{|h|}\,\mu^{ij}_{+}x^{\mu}_{\,\,,i}x^{\nu}_{\,\,,j}\,
\delta\textsl{g}_{\mu\nu}^{+}+
\int_{\Sigma_{-}}\!\!d^{4}\chi\sqrt{|h|}\,\mu^{ij}_{-}x^{\mu}_{\,\,,i}x^{\nu}_{\,\,,j}\,
\delta\textsl{g}_{\mu\nu}^{-}=0$ was considered instead of equation (\ref{pertouli}). Therefore, equations
$\mu^{ij}_{+}K_{ij}^{+}=0$, $\mu^{ij}_{-}K_{ij}^{-}=0$, $\mu^{ij}_{+\,|j}=0$, $\mu^{ij}_{-\,|j}=0$ were derived
instead of equations (\ref{muu matching}), (\ref{mu matching}). From equations $\mu^{ij}_{+\,|j}=0$,
$\mu^{ij}_{-\,|j}=0$, the matching condition (\ref{mu matching}) is derived. However, since in equations
$\mu^{ij}_{+}K_{ij}^{+}=0$, $\mu^{ij}_{-}K_{ij}^{-}=0$ the combination $\lambda^{ij}_{+}+\lambda^{ij}_{-}$
cannot appear, there arises the unnatural situation that the matching conditions contain undetermined
Lagrange multipliers. This cannot be realistic since Lagrange multipliers are supplementary objects and
additionally any physical matching conditions should be well-defined. However for the case
$K_{ij}^{-}=\eta K_{ij}^{+}$ the matching conditions of \cite{Davidson} reduce to the present matching conditions.

Without the use of Lagrange multipliers, we could also proceed with the variation of (\ref{Stotal}) with respect
to the bulk metric at the brane position and get
\begin{eqnarray}
\delta S\big|_{\!b\!r\!a\!n\!e}&=&\frac{1}{2\kappa_{5}^{2}}\int_{\Sigma}d^{4}\chi\,\sqrt{|h|}\,
\big[(K^{\mu\nu}_{+}-K_{+}h^{\mu\nu})+(K^{\mu\nu}_{-}-K_{-}h^{\mu\nu})
+\kappa_{5}^{2}(T^{\mu\nu}-\lambda h^{\mu\nu})-r_{c}G^{\mu\nu}\big]
\delta\textsl{g}_{\mu\nu}\nn\\
&=&\frac{1}{2\kappa_{5}^{2}}\int_{\Sigma}d^{4}\chi\,\sqrt{|h|}\,\big[(K^{ij}_{+}-K_{+}h^{ij})
+(K^{ij}_{-}-K_{-}h^{ij})+\kappa_{5}^{2}(T^{ij}-\lambda h^{ij})-r_{c}G^{ij}\big]
x^{\mu}_{\,\,,i}x^{\nu}_{\,\,,j}\,\delta\textsl{g}_{\mu\nu}\label{alliws}
\end{eqnarray}
If $\delta\textsl{g}_{\mu\nu}$ are independent, we take again the Israel matching conditions. If
$\delta\textsl{g}_{\mu\nu}$ are subject to (\ref{delta x variation}), we take the matching conditions
(\ref{matching1+})-(\ref{matching2}). This method is a straightforward one which convinces us about the
validity of equations (\ref{matching1+})-(\ref{matching2}).

Let us describe in brief, before we continue, another method for deriving the matching conditions
(\ref{matching1+})-(\ref{matching2}). This method was described in section \ref{General arguments and
introduction of the method} and has also been applied in \cite{kof-ira} in the treatment of a
codimension-2 brane, so it is applicable in all codimensions.
The transverse to the brane first derivative of the metric is in general discontinuous between the two sides
of the brane, and therefore, $\mathcal{G}_{\mu\nu}$ contains a distributional piece beyond the regular one,
which is $\textrm{distr}\,\mathcal{G}_{\mu\nu}\!=\!-[(K_{\mu\nu}-Kh_{\mu\nu})_{+}\!+\!(K_{\mu\nu}
-Kh_{\mu\nu})_{-}]\delta(y)$. As usually done when dealing with
distributional sources, the matching conditions are derived by integrating around the singular space.
In \cite{kof-ira} for a codimension-2 brane the corresponding six-dimensional distributional curly terms of
(\ref{totalvariation}) were integrated over the $(r,\theta)$ transverse disc of radius $\epsilon$ in the limit
$\epsilon\rightarrow 0$. For a codimension-3 defect the appropriate integration would occur in a spherical
region $(r,\theta,\phi)$ of radius $\epsilon$ in the limit $\epsilon\rightarrow 0$. Here, the codimension-1
brane is ``sandwiched'' between two ``parallel'' hypersurfaces $H_{\pm}$ each at a distance $\epsilon$ from the
brane. The integration of the distributional curly bracket in (\ref{totalvariation}) gives in the limit
$\epsilon\rightarrow 0$ the first integral in the third line of (\ref{totalvariationother}).
The second integral in the third line of (\ref{totalvariationother}) consists of the usual ``remnant'' terms
of the metric variation and the volume of integration $M$ refers to the space between the hypersurfaces $H_{\pm}$.
Concerning the Gibbons-Hawking term, this should not now be included in the action
(\ref{Stotal}), since already the correct $K^{\mu\nu}\!-\!Kh^{\mu\nu}$ term is present in
(\ref{totalvariationother}), so the inclusion of $K$ would attribute a wrong factor of two. Therefore, the fifth
line of (\ref{totalvariation}) is absent in (\ref{totalvariationother}), as well as the various $K_{ij}$ terms of
(\ref{totalvariation}) disappear in (\ref{totalvariationother}). The characteristic of the absence of the
Gibbons-Hawking term in this treatment of considering the distributional terms in the action is
similar to the fact that in the standard derivation of the Israel matching conditions from the distributional
Einstein equations there is no Gibbons-Hawking term. Finally, the variation (\ref{totalvariation}) gets the
alternative form
\begin{eqnarray}
\delta(S\!+\!S_{c})\big|_{\!b\!r\!a\!n\!e}\!\!&=&\!\!\int_{\!\Sigma_{\pm}}\!\!\!d^{4}\chi \sqrt{|h|}\,
\big(\tilde{\lambda}^{i}x^{\mu}_{\,\,,i}\!+\!2\tilde{\lambda}^{0}n^{\mu}\big)\delta n_{\mu}
+\!\int_{\!\Sigma_{\pm}}\!\!\!d^{4}\chi \sqrt{|h|}\,\tilde{\lambda}^{ij}\delta h_{ij}
+\!\int_{\!\Sigma}d^{4}\chi \sqrt{|h|}\,\Big[\frac{1}{2}(T^{ij}\!-\!\lambda h^{ij})
\!-\!\frac{r_{c}}{2\kappa_{5}^{2}}G^{ij}\!\Big]\delta h_{ij}\nn\\
&-&\!\!\int_{\!\Sigma_{\pm}}\!\!\!d^{4}\chi\,\sqrt{|h|}\,\big(\tilde{\lambda}^{ij}x^{\mu}_{\,\,,i}x^{\nu}_{\,\,,j}
\!+\!\tilde{\lambda}^{0}n^{\mu}n^{\nu}\big) \delta\textsl{g}_{\mu\nu}\nn\\
&+&\!\!\frac{1}{2\kappa _{5}^{2}}\int_{\Sigma_{\pm}}\!\!\!d^{4}\chi \sqrt{|h|}\,\big(K^{\mu\nu}\!-\!Kh^{\mu\nu}\big)
\delta\textsl{g}_{\mu\nu}
\!+\frac{1}{\kappa_{5}^{2}}\int_{\!M}\!d^{5}x\,\sqrt{|\textsl{g}|}\, \textsl{g}^{\lambda[\mu}\,
\textsl{g}^{\kappa]\nu}\,(\delta\textsl{g}_{\mu\nu})_{;\kappa\lambda}\,,
\label{totalvariationother}
\end{eqnarray}
where the various coefficients $\tilde{\lambda}$'s are in general different from $\lambda$'s. Since the
quantities which multiply $(\delta\textsl{g}_{\mu\nu})_{;\kappa\lambda}$ in (\ref{totalvariationother}) do
not have distributional pieces and the variational fields $\delta\textsl{g}_{\mu\nu}$ are considered as
usually smooth functions, the corresponding integral vanishes at the shrink limit and
(\ref{totalvariationother}) becomes
\begin{eqnarray}
\delta(S\!+\!S_{c})\big|_{\!b\!r\!a\!n\!e}\!\!&=&\!\!\int_{\!\Sigma_{\pm}}\!\!\!d^{4}\chi \sqrt{|h|}\,
\big(\tilde{\lambda}^{i}x^{\mu}_{\,\,,i}\!+\!2\tilde{\lambda}^{0}n^{\mu}\big)\delta n_{\mu}
+\!\int_{\!\Sigma_{\pm}}\!\!\!d^{4}\chi \sqrt{|h|}\,\tilde{\lambda}^{ij}\delta h_{ij}
+\!\int_{\!\Sigma}d^{4}\chi \sqrt{|h|}\,\Big[\frac{1}{2}(T^{ij}\!-\!\lambda h^{ij})
\!-\!\frac{r_{c}}{2\kappa_{5}^{2}}G^{ij}\!\Big]\delta h_{ij}\nn\\
&-&\!\!\int_{\!\Sigma_{\pm}}\!\!\!d^{4}\chi\,\sqrt{|h|}\,\big(\tilde{\lambda}^{ij}x^{\mu}_{\,\,,i}x^{\nu}_{\,\,,j}
\!+\!\tilde{\lambda}^{0}n^{\mu}n^{\nu}\big) \delta\textsl{g}_{\mu\nu}
+\frac{1}{2\kappa _{5}^{2}}\int_{\Sigma_{\pm}}\!\!\!d^{4}\chi \sqrt{|h|}\,\big(K^{\mu\nu}\!-\!Kh^{\mu\nu}\big)
\delta\textsl{g}_{\mu\nu}\,,
\label{totalvariationotherfinal}
\end{eqnarray}
There is another way to see why the last integral of (\ref{totalvariationother}) vanishes. This integral,
being a volume integral, can be converted to a surface integral. So, it takes the form of the second integral
of the fourth line in (\ref{totalvariation}) with the only difference that the sign $\Sigma_{\pm}$ of
(\ref{totalvariation}) has to be replaced by
$\Sigma_{\pm}\cup H_{\pm}$. However, the normals $n^{\lambda}$ of the two hypersurfaces $\Sigma_{+}$, $H_{+}$
are opposite to each other, and therefore, at the shrink limit $\epsilon\rightarrow 0$  the quantities
$h^{\mu\nu}n^{\lambda}[(\delta\textsl{g}_{\mu\lambda})_{;\nu}-(\delta\textsl{g}_{\mu\nu})_{;\lambda}]$
from $\Sigma_{+}$, $H_{+}$ cancel each other (the same also happens for $\Sigma_{-}$, $H_{-}$). Continuing the process
from equation (\ref{totalvariationotherfinal}) we arrive anew to the matching conditions
(\ref{matching1+})-(\ref{matching2}). Note that the independent bulk metric variation $\delta\textsl{g}_{\mu\nu}$
in the space outside $H_{\pm}$ gives the bulk field equations (\ref{bulk equations}) under some boundary
variational condition on $H_{\pm}$ (Newmann-like or if the Gibbons-Hawking term is added on $H_{\pm}$
a Dirichlet one).

The general matching conditions (\ref{matching1+}), (\ref{matching1-}) do not provide the equation of motion
for the defect, but only relations on the discontinuity of the extrinsic curvature. However, imposing a relation
between $K_{ij}^{+},K_{ij}^{-}$, an equation of motion arises. E.g. for $K_{ij}^{-}=\eta K_{ij}^{+}$, equation
(\ref{match1}) is the algebraic in the extrinsic curvature equation of motion.
It is a quadratic equation in the extrinsic curvature, contrary to the Israel matching condition
which is linear in the extrinsic curvature.
Equation (\ref{match1}) is the generalization of the Nambu-Goto equation of motion
when the self-gravitating brane interacts with bulk gravity. In the limiting case of no back-reaction,
a probe brane with tension $\lambda$ moving in a fixed background arises. Indeed, in the probe limit, all the
geometric quantities $h_{ij}$, $K_{ij}$, $G_{ij}$ get their background values when the bulk gravity coupling
goes to zero (i.e. $1/\kappa_{5}^{2}\rightarrow 0$) and the extra brane
sources vanish (i.e. $T_{ij}\rightarrow 0$, $r_{c}/\kappa_{5}^{2}\rightarrow 0$). Then,
equation (\ref{match1}) becomes $h^{ij}K_{ij}=0$ which is the Nambu-Goto equation of motion.
Inversely, whenever any extra term beyond $\lambda h^{ij}K_{ij}$ (or all terms) appears
in (\ref{match1}), (\ref{match2}) and these equations are consistent with all the other bulk equations,
then these matching conditions are meaningful back-reacted matching conditions.
In this spirit, the ``smooth'' matching conditions (\ref{match1RT}), (\ref{match2RT}) without extrinsic curvature
discontinuity form an unusual but interesting example. In this case, only the localized
matter and four-dimensional gravity terms participate in the brane equations of motion, and although the
higher-dimensional bulk
terms do not have a direct imprint in these equations, there is still back-reaction since the bulk
equations have also to be satisfied at the brane position. These ``smooth'' matching conditions correspond
to the Regge-Teitelboim equations of motion \cite{Teitelboim}, \cite{battye} with the crucial
difference, however, that there, there are no higher-dimensional gravity terms in the action and the bulk is
prefixed (usually Minkowski). Therefore, possible difficulties discussed in \cite{Deser} are irrelevant here, since
they emanate from the embeddibility restrictions in the given non-dynamical bulk space, while the matching
conditions here dynamically propagate in a non-trivial bulk space.
``Smooth'' matching conditions are also meaningful in codimension-1 standard treatment
\cite{kofinasR4}, without of course the $K_{ij}$ contraction (where there is no balance of
distributional terms between the two sides of the distributional equation, but the right-hand side vanishes on
its own), although there, they lose their significance since there is no Nambu-Goto probe limit so that these
matching conditions to signal a minimal departure from that limit.

Equation (\ref{matching2}) is the second matching condition and expresses a non-conservation equation of
the brane energy-momentum tensor, where the energy exchange between the brane and the bulk is due to the
variability along the brane of the extrinsic geometry. Actually, equation (\ref{matching2}) arises also
in the conventional treatment by differentiation of the standard matching conditions. In the following,
equation (\ref{matching2}) will be written in a more convenient form (also present in the standard approach),
from where it will be seen that the
non-conservation of energy is only due to a possible non-vanishing brane-bulk energy-momentum exchange.

Having finished with the brane equations of motion arising from the distributional parts, we pass to the bulk
equations of motion. These bulk equations are also defined limitingly on the brane, and therefore,
additional equations have to be satisfied at the brane position beyond the matching conditions.
In the Gaussian-normal coordinate system the spacetime metric takes the form
\begin{equation}
ds_{5}^{2}=dy^{2}+g_{ij}(\chi,y)d\chi^{i}d\chi^{j}\,,
\label{metric}
\end{equation}
where the braneworld metric $h_{ij}(\chi)=g_{ij}(\chi,0)$ is assumed to be regular everywhere with the
possible exception of isolated singular points.
From appendix \ref{geometric components}, the $55$ component of the bulk equations
(\ref{bulk equations}) at the brane position gets the form (the index 5 refers to the extra dimension $y$)
\begin{equation}
K_{ij}K^{ij}-K^{2}+R=2\Lambda_{5}-2\kappa_{5}^{2}\mathcal{T}_{55}\,,
\label{goal}
\end{equation}
which holds separately on each side of the brane. Then,
the algebraic matching conditions (\ref{matching1+}), (\ref{matching1-}) are written equivalently in the
following form
\begin{eqnarray}
&&\big[(K^{ij}_{-}-K_{-}h^{ij})+\kappa_{5}^{2}(T^{ij}-\lambda h^{ij})-r_{c}G^{ij}\big]
K_{ij}^{+}=R+2\kappa_{5}^{2}\mathcal{T}_{55}^{+}-2\Lambda_{5}\label{cross1}\\
&&\big[(K^{ij}_{+}-K_{+}h^{ij})+\kappa_{5}^{2}(T^{ij}-\lambda h^{ij})-r_{c}G^{ij}\big]
K_{ij}^{-}=R+2\kappa_{5}^{2}\mathcal{T}_{55}^{-}-2\Lambda_{5}\label{cross2}\,.
\end{eqnarray}
Accordingly, the matching condition (\ref{match1}) for $K_{ij}^{-}=\eta K_{ij}^{+}=\eta K_{ij}$ is written
equivalently in the following simpler form linear in the extrinsic curvature
\begin{equation}
\big[\kappa_{5}^{2}(T^{ij}\!-\!\lambda h^{ij})-r_{c}G^{ij}\big]K_{ij}=(1+\eta)\big(R+2\kappa_{5}^{2}
\mathcal{T}_{55}-2\Lambda_{5}\big)\,,
\label{kala}
\end{equation}
where $\mathcal{T}_{55}\equiv\mathcal{T}_{55}^{+}$,
while in particular the $Z_{2}$-equation (\ref{match1Z}) gets the form
\begin{equation}
\big[\kappa_{5}^{2}(T^{ij}\!-\!\lambda h^{ij})-r_{c}G^{ij}\big]K_{ij}=2\big(R+2\kappa_{5}^{2}
\mathcal{T}_{55}-2\Lambda_{5}\big)\,.
\label{krasia}
\end{equation}

Similarly, from appendix \ref{geometric components}, the $i5$ component of the bulk equations
(\ref{bulk equations}) at the brane position gets the form
\begin{equation}
K^{ij}_{\,\,\,\,|j}-K_{|j}h^{ij}=\kappa_{5}^{2}\mathcal{T}^{\,i5}\,,
\label{paros}
\end{equation}
which holds separately on each side of the brane.
Then, the non-conservation equation (\ref{matching2}) gets a simpler form
\begin{equation}
T^{ij}_{\,\,\,\,|j}=-\big(\mathcal{T}^{\,i5}_{+}+\mathcal{T}^{\,i5}_{-}\big)\,,
\label{equiv}
\end{equation}
which expresses the non-conservation of the brane energy-momentum due to the flux of energy-momentum from the
two sides of the brane. Accordingly, equation (\ref{match2}) is written as
\begin{equation}
T^{ij}_{\,\,\,\,|j}=-(1+\eta)\mathcal{T}^{\,i5}\,,
\label{equiv2}
\end{equation}
where $\mathcal{T}^{\,i5}\equiv\mathcal{T}^{\,i5}_{+}$,
while in particular the $Z_{2}$-equation (\ref{match2Z}) gets the form
\begin{equation}
T^{ij}_{\,\,\,\,|j}=-2\mathcal{T}^{\,i5}\,.
\label{equiv2Z}
\end{equation}

Therefore, for the case that the extrinsic curvatures on the two sides of the brane are linearly related,
the system of equations which have to be satisfied on the brane consists of equations (\ref{match1})
(or equivalently (\ref{kala})), (\ref{match2}) (or equivalently (\ref{equiv2})), (\ref{goal}) and (\ref{paros}).
What remain to be satisfied on the brane are the $ij$ bulk equations which however contain beyond $h_{ij}, K_{ij}$
also $\partial_{y}K_{ij}$, so they are decoupled from the other equations. The previous system of equations for
$K_{ij}$ forms a set of 6 algebraic-differential equations for the 10 independent components of $K_{ij}$,
therefore, the brane-bulk system is always consistent. In the next sections, we are going to study the full system
of brane-bulk equations for the case of cosmology, therefore the brane evolution and the dynamical bulk space
will be found. After the brane geometry is determined, the brane data will be used as initial data to determine the
evolution of the bulk geometry. Concerning this point, the main difference with \cite{Davidson} is that there,
although the brane evolution is also back-reacted, however, the bulk was assumed from the beginning to be the AdS$_{5}$
space.

\section{Codimension-1 brane cosmology with $Z_{2}$-symmetry}
\label{Cosmol}

We focus on the $Z_{2}$-symmetric case and consider that the only bulk energy-momentum content consists
of a pure cosmological constant $\Lambda$. We rewrite the matching conditions (\ref{match1Z}), (\ref{match2Z})
\begin{eqnarray}
&&\Big[K^{ij}-Kh^{ij}+\frac{1}{4M^{3}}(T^{ij}-V h^{ij})-\frac{m^{2}}{2M^{3}}G^{ij}\Big]
K_{ij}=0\label{cosmomatching1Z}\\
&&T^{ij}_{\,\,\,\,|j}=-4M^{3}\big(K^{ij}-Kh^{ij}\big)_{|j}\,,
\label{cosmomatching2Z}
\end{eqnarray}
or the equivalent equations (\ref{krasia}), (\ref{equiv2Z})
\begin{eqnarray}
&&\big(T^{ij}-Vh^{ij}-2m^{2}G^{ij}\big)K_{ij}=4(M^{3}R-\Lambda)
\label{cosmoequiv1Z}\\
&&T^{ij}_{\,\,\,\,|j}=0\,.
\label{cosmoequiv2Z}
\end{eqnarray}

In order to search for cosmological solutions we consider the corresponding form for the bulk metric in the
Gaussian-normal coordinates
\begin{equation}
ds_{5}^{2}=dy^{2}-n^{2}(t,y)dt^{2}+a^{2}(t,y)\,\gamma_{\hat{i}\hat{j}}(\chi^{\hat{\ell}})
d\chi^{\hat{i}}d\chi^{\hat{j}}\,,
\label{bulk metric}
\end{equation}
where $\gamma_{\hat{i}\hat{j}}$ is a maximally symmetric 3-dimensional metric ($\hat{i},\hat{j},...=1,2,3$)
characterized by its spatial curvature $k=-1,0,1$. The energy-momentum tensor on the brane (beyond that of the
brane tension $V>0$) is assumed to be the one of a perfect cosmic fluid with energy density $\rho$ and pressure $p$.

It is convenient to define the quantities
\begin{eqnarray}
A=\frac{a'}{a}\,&,&\,N=\frac{n'}{n}\label{notation1}\\
X=H^{2}+\frac{k}{a^{2}}\,&,&\,Y=\frac{\dot{H}}{n}+H^{2}=\frac{\dot{X}}{2nH}+X
\,\,\,\,\,,\,\,\,\,\,H=\frac{\dot{a}}{na}\,,
\label{notation2}
\end{eqnarray}
where a prime denotes $\partial/\partial{y}$ and a dot denotes $\partial/\partial{t}$.
The cosmic scale factor, lapse function and
Hubble parameter arise as the restrictions on the brane of the functions $a(t,y),n(t,y)$ and $H(t,y)$
respectively. Other quantities also have their corresponding values when restricted on the brane,
and since all the following equations in this section will refer to the brane position, we will use the same
symbols for the restricted quantities without confusion.

The $05$ bulk equation (\ref{paros}) at the position of the brane becomes
\begin{equation}
\dot{A}+nH(A-N)=0\,,
\label{05cosmo}
\end{equation}
while the $55$ bulk equation (\ref{goal}) becomes
\begin{equation}
A(A+N)-(X+Y)+\frac{\Lambda}{6M^{3}}=0\,.
\label{55cosmo}
\end{equation}
Equation (\ref{cosmoequiv1Z}) is linear in $A,N$, and the system (\ref{cosmoequiv1Z}), (\ref{55cosmo}) could
be algebraically solved for $A,N$ in terms of $X,Y$ and replaced into equation (\ref{05cosmo}). However, the arising
equation would contain $\ddot{H}$ and the difficulty for integrating such an equation would increase considerably.
Hopefully, we can do better because $A$ can be integrated from equations (\ref{05cosmo}), (\ref{55cosmo}).
Indeed, eliminating $N$ between equations (\ref{05cosmo}), (\ref{55cosmo}) we obtain the equation
\begin{equation}
(A^{2})^{^{\centerdot}}+4nHA^{2}-2nH\Big(X+Y-\frac{\Lambda}{6M^{3}}\Big)=0\,,
\label{eliminate}
\end{equation}
and since $X+Y=(Xa^{4})^{^{\centerdot}}/(2nHa^{4})$, equation (\ref{eliminate}) becomes a total
derivative
\begin{equation}
\Big(A^{2}a^{4}-Xa^{4}+\frac{\Lambda}{12M^{3}}a^{4}\Big)^{^{\!\centerdot}}=0\,.
\label{totalderiv}
\end{equation}
The integration of (\ref{totalderiv}) gives the equation
\begin{equation}
A^{2}-X+\frac{\Lambda}{12M^{3}}+\frac{\mathcal{C}}{a^{4}}=0
\label{Cintegral}
\end{equation}
($\mathcal{C}$ is integration constant), with two branches for $A$
\begin{equation}
A=\pm\sqrt{X-\frac{\mathcal{C}}{a^{4}}-\frac{\Lambda}{12M^{3}}}\,\,.
\label{Abranches}
\end{equation}

Since $G^{0}_{0}=-3X$, $G^{\hat{i}}_{\hat{j}}=-(X+2Y)\delta^{\hat{i}}_{\hat{j}}$, the matching condition
(\ref{cosmoequiv1Z}) becomes
\begin{equation}
3\big[p-V+2m^{2}(X+2Y)\big]A-\big(\rho+V-6m^{2}X\big)N=24M^{3}(X+Y)-4\Lambda\,,
\label{linearAN}
\end{equation}
from which the quantity $N$ could also be found.
Combining equations (\ref{55cosmo}), (\ref{linearAN}) to eliminate $N$, we obtain the following algebraic
equation for $A$
\begin{equation}
\big(\rho+3p-2V+12m^{2}Y\big)A^{2}-4\big[6M^{3}(X+Y)-\Lambda\big]A-\big(\rho+V-6m^{2}X\big)
\Big(X+Y-\frac{\Lambda}{6M^{3}}\Big)=0\,.
\label{Aalgebraic}
\end{equation}
Substituting $A$ from (\ref{Abranches}) in (\ref{Aalgebraic}), we obtain the final Raychaudhuri equation
for the brane cosmology
\begin{equation}
\Big(\frac{\rho+3p-2V}{4M^{3}}+\frac{3m^{2}}{M^{3}}Y\Big)\Big(X-\frac{\mathcal{C}}{a^{4}}-
\frac{\Lambda}{12M^{3}}\Big)=\Big(X+Y-\frac{\Lambda}{6M^{3}}\Big)\Big(\frac{\rho+V}{4M^{3}}
-\frac{3m^{2}}{2M^{3}}X\pm 6\sqrt{X-\frac{\mathcal{C}}{a^{4}}-\frac{\Lambda}{12M^{3}}}\Big)\,.
\label{roll}
\end{equation}
This equation contains one integration constant, and therefore is distinct from the corresponding equation of
\cite{Davidson}, where there is no integration constant present. In the following, we are going to ignore the
four-dimensional Einstein term \cite{deffayet} and set $m=0$, because in this case we were able to integrate equation
(\ref{roll}). However, before doing so, we can study the limiting case where the 5-dimensional gravity vanishes and only
the 4-dimensional one is present (this corresponds to the Regge-Teitelboim equation of motion \cite{Teitelboim}, but
with the difference that now the bulk is not prefixed). Therefore, setting $M=0$ with $m\neq 0$, $\Lambda\neq 0$ in
equation (\ref{roll}) we get
\begin{equation}
\frac{\dot{H}}{n}+2H^{2}+\frac{k}{a^{2}}=\frac{\rho-3p+4V}{12m^{2}}\,.
\label{ksd}
\end{equation}
For a single component perfect fluid with $p=w\rho$, its conservation equation (\ref{cosmoequiv2Z})
takes the standard form
\begin{equation}
\dot{\rho}+3nH(\rho+p)=0
\label{conserva}
\end{equation}
and (\ref{ksd}) is integrated to
\begin{equation}
H^{2}+\frac{k}{a^{2}}=\frac{V}{6m^{2}}+\frac{\rho}{6m^{2}}+\frac{\mathcal{C}_{1}}{a^{4}}\,,
\label{akk}
\end{equation}
where $\mathcal{C}_{1}$ is integration constant. This is the standard FRW solution with cosmological constant,
but with an extra dark radiation term.

From now on, we set $m=0$ and equation (\ref{roll}) is written as
\begin{equation}
\frac{\dot{H}}{n}+2H^{2}+\frac{k}{a^{2}}-\frac{\Lambda}{6M^{3}}=\frac{\rho+3p-2V}{4M^{3}}
\Big(H^{2}+\frac{k}{a^{2}}-\frac{\mathcal{C}}{a^{4}}-\frac{\Lambda}{12M^{3}}\Big)
\Bigg[\frac{\rho+V}{4M^{3}}\pm6\sqrt{H^{2}+\frac{k}{a^{2}}-\frac{\mathcal{C}}{a^{4}}
-\frac{\Lambda}{12M^{3}}}\,\Bigg]^{-1}\,.
\label{ram}
\end{equation}
It is seen from (\ref{ram}) that for $\mathcal{C}=k=\rho=p=0$, the lower branch contains as solution the Minkowski
brane under the assumption of the Randall-Sundrum fine-tuning $\Lambda+V^{2}/(12M^{3})=0$ \cite{rs}. We will not
assume this condition in our analysis, so in the absence of matter our cosmology may have a de-Sitter vacuum.
It is assumed that the quantity inside the square root in equation (\ref{ram}) is positive.

Equation (\ref{ram}), although pretty complicated, it can be integrated.
We consider a single component perfect fluid with $p=w\rho$ and its conservation equation (\ref{cosmoequiv2Z})
takes the standard form (\ref{conserva}). We can show that for $\Lambda\neq 0$ the variable
\begin{equation}
\Xi=\frac{1}{2}\ln{\Big[\frac{12M^{3}}{|\Lambda|}\Big(H^{2}+\frac{k}{a^{2}}-\frac{\mathcal{C}}{a^{4}}
-\frac{\Lambda}{12M^{3}}\Big)\Big]}
\label{xi}
\end{equation}
obeys the differential equation
\begin{equation}
\frac{d\Xi}{d\tilde{\rho}}=\frac{1}{3(1\!+\!w)(\tilde{\rho}-\tilde{V})}\Big[2-\frac{(1\!+\!3w)\tilde{\rho}-
3(1\!+\!w)\tilde{V}}{\tilde{\rho}\pm 6e^{\Xi}}\Big]\,,
\label{xiequation}
\end{equation}
where
\begin{equation}
\tilde{\rho}=\sqrt{\frac{12M^{3}}{|\Lambda|}}\,\frac{\rho+V}{4M^{3}}=\frac{\rho+V}{\rho_{\ast}}
\quad\quad,\quad\quad\tilde{V}=\frac{V}{\rho_{\ast}}\quad\quad,\quad\quad
\rho_{\ast}=4M^{3}\sqrt{\frac{|\Lambda|}{12M^{3}}}\,.\label{tilde}
\end{equation}
Note that the Randall-Sundrum fine-tuning corresponds to the value $\tilde{V}=3$. Finally, changing to the
variable
\begin{equation}
\Phi=(\tilde{\rho}\pm 6 e^{\Xi})^{2}\,,
\label{Phi}
\end{equation}
we get, after some cancelations, a linear differential equation
\begin{equation}
\frac{d\Phi}{d\tilde{\rho}}-\frac{4}{3(1\!+\!w)(\tilde{\rho}-\tilde{V})}\Phi=2\tilde{\rho}\,
\frac{(1\!+\!3w)\tilde{\rho}-3(1\!+\!w)\tilde{V}}{3(1\!+\!w)(\tilde{\rho}-\tilde{V})}\,,
\label{ode}
\end{equation}
with general solution
\begin{equation}
\Phi=\Big(\frac{\rho}{\rho_{\ast}}+\tilde{V}\Big)^{2}+c\,\Big(\frac{\rho}{\rho_{\ast}}\Big)^{\frac{4}{3(1+w)}}\,,
\label{soloPhi}
\end{equation}
where $c$ is integration constant. For $c<0$, due to that $\Phi>0$, there are some restrictions on the
allowed values of $\rho$.

From the definition (\ref{Phi}) we can find that
\begin{equation}
\tilde{\rho}\pm \frac{24M^{3}}{\rho_{\ast}}
\sqrt{H^{2}+\frac{k}{a^{2}}-\frac{\mathcal{C}}{a^{4}}-\frac{\Lambda}{12M^{3}}}
=\epsilon\,\sqrt{\Phi}\,.
\label{sqrtPhi}
\end{equation}
In this equation the sign index $\epsilon=+1$ or $-1$ has been used to denote a new different bifurcation
from the previous $\pm$ branches. It is seen from (\ref{sqrtPhi}) that the sign $\epsilon=-1$ is only
consistent with the lower $\pm$ branch, while the sign $\epsilon=+1$ is consistent with both $\pm$ branches.
The distinction, however, introduced by the sign index $\pm$ will be lost
in the following expressions for the expansion rate and the acceleration parameter and only the sign $\epsilon$
will distinguish the two branches of solutions.
\newline
{\textit{The expansion rate}} of the new cosmology {\textit{for $\Lambda\neq 0$}} arises by squaring equation
(\ref{sqrtPhi}) and is given by
\begin{equation}
{\underline{H^{2}+\frac{k}{a^{2}}-\frac{\mathcal{C}}{a^{4}}=\Big(\frac{\rho_{\ast}}{24M^{3}}\Big)^{\!2}\,
\Bigg{\{}\Bigg[\frac{\rho}{\rho_{\ast}}+\tilde{V}-\epsilon\,\sqrt{\Big(\frac{\rho}{\rho_{\ast}}
+\tilde{V}\Big)^{\!2}+c\,\Big(\frac{\rho}{\rho_{\ast}}\Big)^{\!\frac{4}{3(1+w)}}}\,\Bigg]^{2}\!+36\,
\text{sgn}({\Lambda})\Bigg{\}}}}\,,
\label{H}
\end{equation}
where $\text{sgn}({\Lambda})$ is the sign of $\Lambda$. The positiveness of the quantity inside the square
root of equation (\ref{ram}) is now always assured.
The solution (\ref{H}) contains two integrations constants. The first constant $\mathcal{C}$ is associated with the usual
dark radiation term reflecting the non-vanishing bulk Weyl tensor. The second constant $c$ is the new feature
that does not appear in the cosmology of the standard matching conditions \cite{binetruy} and signals new
characteristics in the cosmic evolution. Setting $c=0$ in the branch $\epsilon=-1$ we obtain the cosmology
of the standard matching conditions $H^{2}+\frac{k}{a^{2}}-\frac{\mathcal{C}}{a^{4}}=
\big(\frac{\rho+V}{12M^{3}}\big)^{\!2}+\frac{\Lambda}{12M^{3}}$. Of course, there is always the extra integration
constant from the integration of the conservation equation (\ref{conserva}) for $\rho$ which is adjusted by the
today matter content. We will see that $c$ will be accounted for this today matter content (or equivalently the
today dark energy content). The solution also contains three free parameters $M$, $\Lambda$, $V$
or $M$, $\rho_{\ast}$, $\tilde{V}$.

For $\Lambda=0$ the variable
\begin{equation}
\bar{\Xi}=\frac{1}{2}\ln{\Big[\frac{16M^{6}}{V^{2}}\Big(H^{2}+\frac{k}{a^{2}}-\frac{\mathcal{C}}{a^{4}}\Big)\Big]}
\label{xi0}
\end{equation}
obeys the differential equation
\begin{equation}
\frac{d\bar{\Xi}}{d\bar{\rho}}=\frac{1}{3(1\!+\!w)(\bar{\rho}-1)}\Big[2-\frac{(1\!+\!3w)\bar{\rho}-
3(1\!+\!w)}{\bar{\rho}\pm 6e^{\bar{\Xi}}}\Big]\,,
\label{xi0equation}
\end{equation}
where
\begin{equation}
\bar{\rho}=\frac{\rho}{V}+1\,.\label{bar}
\end{equation}
Defining
\begin{equation}
\bar{\Phi}=(\bar{\rho}\pm 6 e^{\bar{\Xi}})^{2}\,,
\label{Phi0}
\end{equation}
we get the equation
\begin{equation}
\frac{d\bar{\Phi}}{d\bar{\rho}}-\frac{4}{3(1\!+\!w)(\bar{\rho}-1)}\bar{\Phi}=2\bar{\rho}\,
\frac{(1\!+\!3w)\bar{\rho}-3(1\!+\!w)}{3(1\!+\!w)(\bar{\rho}-1)}\,,
\label{0ode}
\end{equation}
with general solution
\begin{equation}
\bar{\Phi}=\Big(\frac{\rho}{V}+1\Big)^{2}+\bar{c}\,\Big(\frac{\rho}{V}\Big)^{\frac{4}{3(1+w)}}\,,
\label{0soloPhi}
\end{equation}
where $\bar{c}$ is integration constant. Finally, the {\textit{expansion rate for $\Lambda=0$}} is
\begin{equation}
H^{2}+\frac{k}{a^{2}}-\frac{\mathcal{C}}{a^{4}}=\Big(\frac{V}{24M^{3}}\Big)^{\!2}\,
\Bigg[\frac{\rho}{V}+1-\epsilon\,\sqrt{\Big(\frac{\rho}{V}
+1\Big)^{\!2}+\bar{c}\,\Big(\frac{\rho}{V}\Big)^{\!\frac{4}{3(1+w)}}}\,\Bigg]^{2}\,.
\label{H0}
\end{equation}

\section{Bulk solution}
\label{bulk sol}

Equations (\ref{05cosmo})-(\ref{totalderiv}) are not only valid on the brane, but also in the bulk.
The $00$ bulk equation (\ref{bulk equations}) is written as
\begin{equation}
A'+2A^{2}-X+\frac{\Lambda}{6M^{3}}=0.
\label{00bulk}
\end{equation}
From equation (\ref{05cosmo}) we get $H'=-HA$, $X'=-2XA$ and using (\ref{00bulk}) we obtain
\begin{equation}
\Big(A^{2}a^{4}-Xa^{4}+\frac{\Lambda}{12M^{3}}a^{4}\Big)'=0\,.
\label{totalprime}
\end{equation}
From (\ref{totalderiv}), (\ref{totalprime}), we obtain the integral (\ref{Cintegral}), but now valid
everywhere in the bulk, with $\mathcal{C}$ integration constant. Finally, the spatial $\hat{i}\hat{j}$ bulk
equations (\ref{bulk equations}) consist of only one equation
\begin{equation}
2A'+N'+3A^{2}+N^{2}+2AN-X-2Y+\frac{\Lambda}{2M^{3}}=0\,.
\label{stoa}
\end{equation}
From (\ref{05cosmo}) we get $\frac{1}{n}\dot{H}'=H^{2}(2A\!-\!N)\!-\!AY$, $Y'=-(A\!+\!N)Y$.
Then, differentiating (\ref{55cosmo}) with respect to $y$ and using (\ref{00bulk}) we  get
an equation for $N'$
\begin{equation}
AN'=(2A+N)\Big(2A^{2}-X+\frac{\Lambda}{6M^{3}}\Big)-2AX-(A+N)Y\,.
\label{Nprime}
\end{equation}
Substituting equations (\ref{00bulk}), (\ref{Nprime}) into (\ref{stoa}) and using (\ref{55cosmo}),
equation (\ref{stoa}) becomes identically satisfied.

Equation (\ref{05cosmo}) is also written as
\begin{equation}
\Big(\frac{\dot{a}}{n}\Big)'=0\,,
\label{ger}
\end{equation}
which is integrated to
\begin{equation}
\dot{a}=f(t)\,n\,,
\label{inte}
\end{equation}
where $f(t)$ is an arbitrary function of time. Adopting the cosmic time on the brane $n(t,0)=1$ we have
$f(t)=\dot{a}_{0}$, where we adopt temporarily in this section the notation that $a_{0}$ denotes the brane scale
factor. It is
\begin{equation}
n(t,y)=\frac{\dot{a}(t,y)}{\dot{a}_{0}(t)}\,.
\label{nty}
\end{equation}
Using (\ref{nty}), equation (\ref{totalprime}) becomes
\begin{equation}
(a^{2})''+\frac{\Lambda}{3M^{3}}a^{2}=g(t)\,,
\label{rola}
\end{equation}
where $g(t)=2(\dot{a}_{0}^{2}+k)$. The solution of (\ref{rola}) is
\begin{eqnarray}
a^{2}(t,y)=\Bigg\{\!\!\begin{array}{c}\mathcal{A}_{1}\cosh{(\mu y)}+\mathcal{A}_{2}\sinh{(\mu y)}
-2\mu^{-2}a_{0}^{2}X\,\,\,\,,\,\,\,\,\Lambda<0\\
\mathcal{B}_{1}\cos{(\mu y)}+\mathcal{B}_{2}\sin{(\mu y)}+2\mu^{-2}a_{0}^{2}X\,\,\,\,,\,\,\,\,\Lambda>0
\\ a_{0}^{2}Xy^{2}+\mathcal{D}_{2}y+\mathcal{D}_{1}\,\,\,\,,\,\,\,\,\Lambda=0\end{array}
\,\,\,\,\,\,\,,\,\,\,\,\,\,\,\,\,\,\mu=\sqrt{\frac{|\Lambda|}{3M^{3}}}\,,
\label{nibi}
\end{eqnarray}
where $X$ is given by the brane expansion rate (\ref{H}), (\ref{H0}) and $\mathcal{A}_{1},\mathcal{A}_{2},\mathcal{B}_{1},
\mathcal{B}_{2},\mathcal{D}_{1},\mathcal{D}_{2}$ are functions
of time $t$. This solution corresponds to the right hand side of the brane. Due to the $Z_{2}-$symmetry,
the solution on the left hand side arises by the substitution $\mathcal{A}_{2}\rightarrow -\mathcal{A}_{2}$,
$\mathcal{B}_{2}\rightarrow -\mathcal{B}_{2}$,
$\mathcal{D}_{2}\rightarrow -\mathcal{D}_{2}$. Solution (\ref{nibi}) gives for $y=0$ the coefficients
$\mathcal{A}_{1}=a_{0}^{2}(1+2\mu^{-2}X)$, $\mathcal{B}_{1}=a_{0}^{2}(1-2\mu^{-2}X)$, $\mathcal{D}_{1}=a_{0}^{2}$.
The remaining coefficients are found by differentiating (\ref{nibi}) with respect to $y$ at
the brane position giving $\mathcal{A}_{2}=\mathcal{B}_{2}=2\mu^{-1}a_{0}^{2}A$, $\mathcal{D}_{2}=2a_{0}^{2}A$,
where the brane value $A$ is given by (\ref{Abranches}) in terms of $X$. Finally, we write the bulk solution as
\begin{equation}
\frac{a^{2}(t,y)}{a^{2}(t,0)}=\Bigg\{\!\!\begin{array}{c}(1\!+\!2\mu^{-2}X)\cosh{(\mu y)}+2\mu^{-1}A\sinh{(\mu y)}
-2\mu^{-2}X\,\,\,\,,\,\,\,\,\Lambda<0\\
(1\!-\!2\mu^{-2}X)\cos{(\mu y)}+2\mu^{-1}A\sin{(\mu y)}+2\mu^{-2}X\,\,\,\,,\,\,\,\,\Lambda>0
\\ Xy^{2}+2Ay+1\,\,\,\,,\,\,\,\,\Lambda=0\end{array}\,\,\,\,\,\,\,,\,\,\,\,\,\,\,\,\,\,
\mu=\sqrt{\frac{|\Lambda|}{3M^{3}}}\,,
\label{nibir}
\end{equation}
where $X(t)$ and $A(t)$ are given by (\ref{H}), (\ref{H0}) and (\ref{Abranches}) respectively. We note that the cosmological solution
(\ref{H}) we have derived is quite different than the solution found in \cite{Davidson}, where the domain wall was
assumed to move in a 5-dimensional AdS space. This mismatch shows that the bulk space described by
the solution (\ref{nibir}) is not AdS, but reduces to AdS for the particular choice of the integration constant $c=0$.

\section{Investigation of the Cosmology}
\label{inve}

We will focus in the following in the case of a negative bulk cosmological constant, $\Lambda<0$, which
involves the situation with $\Lambda$ being a small deformation from the Randall-Sundrum value.
The scale factor for the branch $\epsilon=-1$ with $\tilde{V}< 3$
is bounded from above, and the same happens for the branch $\epsilon=+1$ with any value of $\tilde{V}$.
However, the branch $\epsilon=-1$ with $\tilde{V}\geq 3$ possesses the late-times asymptotic linearized regime
with a positive effective cosmological constant
\begin{equation}
H^{2}+\frac{k}{a^{2}}-\frac{\mathcal{C}}{a^{4}}\approx 2\gamma\rho+\frac{\Lambda_{e\!f\!f}}{3}\,,
\label{linear1}
\end{equation}
where
\begin{eqnarray}
&&\gamma=\frac{4\pi G_{N}}{3}=\frac{V}{144M^{6}}\label{gamma}\\
&&\Lambda_{e\!f\!f}=3{\Big({\frac{\rho_{\ast}}{4{M^3}}}\Big)^{\!2}}\,\Big({\frac{\tilde{V}^2}{9}-1}\Big)
=\frac{1}{4M^3}\Big({\Lambda+\frac{V^2}{12 M^3}}\Big)\label{Lamdaeff}\,.
\end{eqnarray}
The identification of Newton's constant $G_{N}$ with parameters of the theory reduces the number of free
parameters from three to two.

At early times, the dominant behaviour for the branch $\epsilon=-1$ is $H^{2}+\frac{k}{a^{2}}-
\frac{\mathcal{C}}{a^{4}}\approx (\frac{\rho}{12M^{3}})^{2}$, while for $\epsilon=+1$ it is
$H^{2}+\frac{k}{a^{2}}-\frac{\mathcal{C}}{a^{4}}\approx \text{constant}$ (which may help to inflation).

In the temporal gauge choice of the brane $n(t,0)=1$ (cosmic time on the brane), the acceleration-deceleration
eras during the cosmic evolution are studied by evaluating the quantity $\ddot{a}/a=\dot{H}+H^{2}$. From
equations (\ref{ram}), (\ref{sqrtPhi}) we obtain
\begin{equation}
\frac{\ddot{a}}{a}=\Big(\frac{\rho+3p-2V}{\epsilon\rho_{\ast}\sqrt{\Phi}}-1\Big)
\Big(\frac{\rho+V-\epsilon\rho_{\ast}\sqrt{\Phi}}{24M^{3}}\Big)^{\!2}+
\frac{\Lambda}{12M^{3}}-\frac{\mathcal{C}}{a^{4}}\,,
\label{accel}
\end{equation}
where $\Phi$ is given by (\ref{soloPhi}). Note that for $\epsilon=-1$, the late-times behaviour
$\rho\rightarrow 0$ of equation (\ref{accel}) is $\ddot{a}/a\rightarrow (\rho_{\ast}/4M^{3})^{2}
\,(\tilde{V}^{2}/9-1)$, which is positive for $\tilde{V}>3$.

It is convenient for the investigation of the cosmological behaviour, instead of using the parameter $\rho$, to
express the expansion rate and the acceleration in terms of the parameter
\begin{equation}
\textrm{y}=\frac{\rho}{\rho_{o}}=(1+z)^{3(1+w)}\,,
\label{y}
\end{equation}
where a subscript $o$ characterizes present values. Today epoch corresponds to $\textrm{y}=1$ and past to
$\textrm{y}>1$. The redshift is denoted as usually by $z$. Then, equations (\ref{H}), (\ref{accel}) become
\begin{equation}
H^{2}+\frac{k}{a^{2}}-\frac{\mathcal{C}}{a^{4}}=\Big(\frac{\rho_{\ast}\tilde{V}}{24M^{3}}\Big)^{\!2}\,
\Big[\Big(1+\frac{x_{o}}{\tilde{V}}\textrm{y}-\epsilon\sqrt{\varphi}\Big)^{\!2}-\frac{36}{\tilde{V}^{2}}\Big]\,,
\label{h}
\end{equation}
\begin{equation}
\frac{\ddot{a}}{a}=\Big(\frac{\rho_{\ast}\tilde{V}}{24M^{3}}\Big)^{\!2}\,
\Bigg{\{}\Bigg[\frac{(1\!+\!3w)(1+\frac{x_{o}}{\tilde{V}}\textrm{y})-3(1\!+\!w)}{\epsilon\,\sqrt{\varphi}}-1\Bigg]
\Big(1+\frac{x_{o}}{\tilde{V}}\textrm{y}-\epsilon\sqrt{\varphi}\Big)^{\!2}
-\frac{36}{\tilde{V}^{2}}\Bigg{\}}-\frac{\mathcal{C}}{a^{4}}\,,
\label{aca}
\end{equation}
where
\begin{equation}
\varphi=\frac{\Phi}{\tilde{V}^{2}}=\Big(1+\frac{x_{o}}{\tilde{V}}\textrm{y}\Big)^{\!2}+\frac{cx_{o}^{4/3}}{\tilde{V}^{2}}
\Big(\frac{\textrm{y}}{x_{o}^{w}}\Big)^{\frac{4}{3(1+w)}}\,\,\,\,\,,\,\,\,\,\,
x_{o}=\frac{\rho_{o}}{\rho_{\ast}}\,.
\label{varphi}
\end{equation}
For $c<0$, due to that $\varphi>0$, there are some restrictions on the allowed values of $\textrm{y}$.

The age of the universe $t_{o}$ can also be expressed in terms of the parameter $\textrm{y}$.
Considering that the main contribution in $t_{o}$ comes from the dust era, we have the approximation
\begin{equation}
t_{o}\approx\frac{1}{3}\int_{1}^{\infty}\!\!\frac{d\textrm{y}}{\textrm{y}H(\textrm{y})}\,.
\label{age}
\end{equation}

We will restrict our attention to the study of the more reasonable branch $\epsilon=-1$ which possesses the
late-times linearized LFRW regime. In the following we provide a more detailed analysis of the phenomenological
consequences of this cosmology. The branch $\epsilon=+1$ is not necessarily precluded, but along with a proper
dark energy/vacuum content \cite{darkreview}, may meet recent observations of the early epoch or of late times.

\subsection{Branch $\epsilon=-1$}

Equation (\ref{linear1}), beyond the linear term, possesses a series of extra terms which become significant
away from the asymptotic regime. These extra terms consist the dark energy contribution, while the linear
term consists the matter contribution observed today. Therefore, the existence of the linear term in the
full expansion assures for $\tilde{V}\geq 3$ the relation (\ref{gamma}) between the parameters of the theory
and Newton's constant. This relation reduces the number of free parameters by one and makes our analytical
estimates relatively easier. In the following analysis we will consider this case $\tilde{V}\geq 3$.

Every contribution in the expansion rate (\ref{H}) can be parameterized by normalized variables
\begin{equation}
\Omega_{m}+\Omega_{DE}+\Omega_{k}+\Omega_{\mathcal{C}}=1\,,
\label{flat}
\end{equation}
where the matter contribution is
\begin{equation}
\Omega_{m}=\frac{2\gamma\rho}{H^{2}}\,,
\label{Omegam}
\end{equation}
the dark energy component is
\begin{equation}
\Omega_{DE}=\Big(\frac{\rho_{\ast}}{24M^{3}}\Big)^{\!2}\frac{1}{H^{2}}
\Bigg{\{}\Bigg[\frac{\rho}{\rho_{\ast}}+\tilde{V}+\sqrt{\Big(\frac{\rho}{\rho_{\ast}}
+\tilde{V}\Big)^{\!2}+c\,\Big(\frac{\rho}{\rho_{\ast}}\Big)^{\!\frac{4}{3(1+w)}}}\,\Bigg]^{2}\!-36-
8\tilde{V}\frac{\rho}{\rho_{\ast}}\Bigg{\}}\,,
\label{OmegaDE}
\end{equation}
while the topology contribution and the dark radiation portion are
\begin{equation}
\Omega_{k}=-\frac{k}{a^{2}H^{2}}\,\,\,\,\,,\,\,\,\,\,\Omega_{\mathcal{C}}=\frac{\mathcal{C}}{a^{4}H^{2}}\,.
\label{Omegak}
\end{equation}
\newline
In order for the energy density of dark radiation $\rho_{\mathcal{C}}=\frac{\mathcal{C}}{2\gamma a^{4}}$ not
to violate the nucleosynthesis constraints it should be $-1.23\leq\frac{\rho_{\mathcal{C}}}{\rho_{r}}\leq 0.11$
\cite{nucl}, where $\rho_{r}$ is the energy density of the radiation component. Since $\rho_{r}$ is insignificant
today, $\rho_{\mathcal{C}}$ is also insignificant today. Therefore, the term $\mathcal{C}/a^{4}$ can be neglected
(at least for a relatively recent epoch).
\newline
Note also that
\begin{equation}
\frac{x_{o}}{\tilde{V}}=\frac{\Omega_{m,o}}{2}\Big(\frac{H_{o}}{12M^{3}\gamma}\Big)^{2}\equiv\eta
\approx 2\times 10^{-30}[M(\text{TeV})]^{-6}\,.
\label{eta}
\end{equation}
Equation (\ref{h}) becomes
\begin{equation}
\frac{1}{(6M^{3}\gamma)^{2}}\Big(H^{2}+\frac{k}{a^{2}}-\frac{\mathcal{C}}{a^{4}}\Big)=\big(1+\sqrt{\varphi}
+\eta\,\textrm{y}\big)^{\!2}-\frac{36}{\tilde{V}^{2}}\,,
\label{gala}
\end{equation}
where
\begin{equation}
\varphi=\big(1+\eta\textrm{y}\big)^{\!2}+\frac{cx_{o}^{4/3}}{\tilde{V}^{2}}
\Big(\frac{\textrm{y}}{x_{o}^{w}}\Big)^{\frac{4}{3(1+w)}}\,.
\label{bgu}
\end{equation}

From equation (\ref{OmegaDE}) it is seen that the integration constant $c$ can be adjusted to get the measured
dark energy component and this is the main difference of the current cosmological model compared to the cosmology
of the standard matching conditions. Combining the present values of equations (\ref{Omegam}), (\ref{OmegaDE})
we get the equation
\begin{equation}
\Big[\tilde{V}+x_{o}+\sqrt{\big(\tilde{V}+x_{o}\big)^{\!2}+c\,x_{o}^{4/3}}\,\Big]^{2}
=4\big(9+2\omega_{o}\tilde{V}x_{o}\big)\,\,\,\,\,\,,\,\,\,\,\,\,\omega_{o}=1+\frac{\Omega_{DE,o}}{\Omega_{m,o}}\,,
\label{c}
\end{equation}
which can be solved for $c$ in terms of the parameters
\begin{equation}
c\,x_{o}^{4/3}
=4\sqrt{9+2\omega_{o}\tilde{V}x_{o}}\,\,\Big[\sqrt{9+2\omega_{o}\tilde{V}x_{o}}
\,-\big(\tilde{V}+x_{o}\big)\Big]\,.
\label{sheep}
\end{equation}
From equations (\ref{c}), (\ref{sheep}) it arises that the parameters have to satisfy the constraint
$(\tilde{V}+x_{o})^{2}<4(9+2\omega_{o}\tilde{V}x_{o})$, or equivalently due to
(\ref{eta}), $\tilde{V}<6+6\big(4\omega_{o}\!-\!1\big)\eta$. Therefore, $x_{o}\lesssim 6\eta$, i.e.
$x_{o}\lesssim 10^{-29}[M(\text{TeV})]^{-6}$. Additionally, since $\tilde{V}x_{o}\lesssim 36\eta$, the quantity
$cx_{o}^{4/3}$ in (\ref{sheep}) can be approximated by
\begin{equation}
\frac{1}{12}c\,x_{o}^{4/3}\approx 3-\tilde{V}+\Big(\frac{\omega_{o}}{3}\tilde{V}-1\Big)\tilde{V}\eta\,,
\label{molibi}
\end{equation}
and set into (\ref{bgu}). Therefore, $c$ is accounted for the today matter content $\Omega_{m,o}$ and the today
dark energy content $\Omega_{DE,o}$ (which are approximately known).

We are now in position to capture the basic cosmological characteristics.
\newline
{\underline{\textit{Dust era}}} ($w=0$): Equation (\ref{aca}), neglecting the dark radiation, becomes
\begin{equation}
\frac{1}{(6M^{3}\gamma)^{2}}\frac{\ddot{a}}{a}=\frac{1}{\sqrt{\varphi}}\big(2-\sqrt{\varphi}-
\eta\,\textrm{y}\big)\big(1+\sqrt{\varphi}+\eta\,\textrm{y}\big)^{\!2}-\frac{36}{\tilde{V}^{2}}\,,
\label{xara}
\end{equation}
where
\begin{equation}
\varphi=(1+\eta\,\textrm{y})^{2}+\frac{cx_{o}^{4/3}}{\tilde{V}^{2}}\textrm{y}^{4/3}\,.
\label{tan}
\end{equation}
We finally observe that equations (\ref{gala}), (\ref{xara}) contain only the two free parameters $\tilde{V}$,
$M$ and the flatness value $\Omega_{m,o}$\,.
\newline
To continue our analysis we discern the following possible cases:
\newline
$\bullet$\,\,\,$\tilde{V}-3>>\eta$ and $6-\tilde{V}>>\sqrt{\eta}$. This case means that $\tilde{V}$ is between
the values 3 and 6 but not extremely close to 3 or 6.
Then, $cx_{o}^{4/3}\approx -12(\tilde{V}-3)$,
$\varphi_{o}\approx \big(\frac{6}{\tilde{V}}-1\big)^{2}$,
$\frac{1}{(6M^{3}\gamma)^{2}}\frac{\ddot{a}}{a}\!\!\mid_{o}\approx\frac{144}{\tilde{V}^{2}}
\frac{\tilde{V}-3}{6-\tilde{V}}>0$. For $\eta\textrm{y}<<1$, i.e. $z<<10^{10}[M(\text{TeV})]^{2}$, it is
$\varphi\approx 1-12\frac{\tilde{V}-3}{\tilde{V}^{2}}\textrm{y}^{4/3}\approx
1-12\frac{\tilde{V}-3}{\tilde{V}^{2}}(1+z)^{4}$ and the requirement $\varphi>0$ means
$z<(\frac{1}{12}\frac{\tilde{V}^{2}}{\tilde{V}-3})^{1/4}-1$. Therefore, except if $\tilde{V}$ is very close
to the value 3, the values of $z$ are restricted close to 0. Furthermore,
$\frac{1}{(6M^{3}\gamma)^{2}}\frac{\ddot{a}}{a}\approx \frac{2}{\sqrt{\varphi}}-\varphi-1
+4(1-\frac{9}{\tilde{V}^{2}})$, which is permanently positive giving always acceleration. So,
these values of $\tilde{V}$ are not particularly interesting for the dust era of the universe.
\newline
$\bullet$\,\,\,$\tilde{V}=6+\textrm{v}$, where $|\textrm{v}|\lesssim\mathcal{O}(\sqrt{\eta})$. This case means
that $\tilde{V}$ is extremely close to the value 6 from above or from below. It is $cx_{o}^{4/3}/12\approx
-3-\textrm{v}+6(2\omega_{o}-1)\eta$, $\varphi_{o}\approx 4\omega_{o}\eta+\frac{\textrm{v}^{2}}{9}$,
$\frac{1}{(6M^{3}\gamma)^{2}}\frac{\ddot{a}}{a}\!\!\mid_{o}\approx\frac{6}{\sqrt{36\omega_{o}\eta+\textrm{v}^{2}}}>0$.
For $\eta\textrm{y}^{4/3}<<1$, i.e.
$z<<10^{7}[M(\text{TeV})]^{3/2}$, it is $\varphi\approx 1-\textrm{y}^{4/3}+2\eta\textrm{y}+2(2\omega_{o}-1)
\eta\textrm{y}^{4/3}+\frac{\textrm{v}^{2}}{9}\textrm{y}^{4/3}$. Therefore, not only the today acceleration is huge, but in order to be
$\varphi>0$, $\textrm{y}$ must be very close to 1, which means that practically there is not past. So,
these values of $\tilde{V}$ do not have particular significance.
\newline
$\bullet$\,\,\,$\tilde{V}=3+v$, where $v=\frac{3}{4}\vartheta\,\omega_{_{\!\!\!\!\!\thicksim}o}\eta$,
$\frac{1}{2}<\vartheta<1-\frac{2}{\omega_{_{\!\!\!\!\!\thicksim}o}}$,
$\omega_{_{\!\!\!\!\!\thicksim}o}=2\big(2\omega_{o}-1\big)
=2\big(1+2\frac{\Omega_{DE,o}}{\Omega_{m,o}}\big)$. It is indeed
$1-\frac{2}{\omega_{_{\!\!\!\!\!\thicksim}o}}>\frac{1}{2}$ if $\Omega_{DE,o}>2\Omega_{m,o}$.
This case corresponds to an extreme fine-tuning close to the Randall-Sundrum
value $\tilde{V}=3$, which, however, possesses interesting phenomenological implications. For the typical
values $\Omega_{m,o}=0.3$, $\Omega_{DE,o}=0.7$, we get $1/2<\vartheta\lesssim 0.82$.
It is $cx_{o}^{4/3}\approx 9\big[(1-\vartheta)\omega_{_{\!\!\!\!\!\thicksim}o}-2\big]
\eta$, $\varphi_{o}\approx 1+(1\!-\!\vartheta)\,
\omega_{_{\!\!\!\!\!\thicksim}o}\eta$ and $\frac{1}{(6M^{3}\gamma)^{2}}\frac{\ddot{a}}{a}\!\!\mid_{o}\approx
2(2\vartheta\!-\!1)\,\omega_{_{\!\!\!\!\!\thicksim}o}\eta>0$, therefore we get acceleration today.
For $\eta\textrm{y}^{4/3}<<1$, i.e.
$z<<10^{7}[M(\text{TeV})]^{3/2}$, it is $\varphi\approx 1+2\eta\textrm{y}+2\big(\frac{1-\vartheta}{2}
\omega_{_{\!\!\!\!\!\thicksim}o}\!-\!1\big)\eta\textrm{y}^{4/3}>0$ and
$\frac{1}{4(6M^{3}\gamma)^{2}}\frac{\ddot{a}}{a}\approx
\big[\frac{\vartheta}{2}\omega_{_{\!\!\!\!\!\thicksim}o}-\textrm{y}-\big(\frac{1-\vartheta}{2}
\omega_{_{\!\!\!\!\!\thicksim}o}\!-\!1\big)\textrm{y}^{4/3}\big]\eta$. Obviously, for $\textrm{y}\rightarrow
+\infty$, it is $\frac{\ddot{a}}{a}\rightarrow -\infty$. Therefore, we have in the past a long deceleration
era resulting to a small acceleration today \cite{Riess}, \cite{WMAP1}, \cite{BAO1}. For the previous
values of $\Omega_{m,o}$, $\Omega_{DE,o}$ the passage from deceleration to acceleration occurs, e.g. at
redshift $z_{p}=0.55$ for $\vartheta=0.8$, or $z_{p}=0.28$ for $\vartheta=0.7$,
or $z_{p}=0.12$ for $\vartheta=0.6$, or $z_{p}=0.01$ for $\vartheta=0.51$ (for the $\Lambda$CDM model the
corresponding passage occurs at redshift $z_{p}=0.67$). The Hubble parameter is
found to be $\frac{1}{(12M^{3}\gamma)^{2}}\big(H^{2}+\frac{k}{a^{2}}-\frac{\mathcal{C}}{a^{4}}\big)
\approx\big[\frac{\vartheta}{2}\omega_{_{\!\!\!\!\!\thicksim}o}+2\textrm{y}+\big(\frac{1-\vartheta}{2}
\omega_{_{\!\!\!\!\!\thicksim}o}\!-\!1\big)\textrm{y}^{4/3}\big]\eta$ and the right-hand side is always
positive. The age of the universe for $k=\mathcal{C}=0$ is given by the expression $t_{o}
\approx \frac{\sqrt{2}}{3H_{o}\sqrt{\Omega_{m,o}}}\int_{1}^{\infty}\!\!\frac{d\textrm{y}}{\textrm{y}}
\big{\{}\frac{\vartheta}{2}\omega_{_{\!\!\!\!\!\thicksim}o}+2\textrm{y}+\big(\frac{1-\vartheta}{2}
\omega_{_{\!\!\!\!\!\thicksim}o}\!-\!1\big)\textrm{y}^{4/3}\big{\}}^{-1/2}$, which e.g. for $\vartheta=0.8$,
$\Omega_{m,o}=0.3$ gives $t_{o}= \frac{8.9}{h}\text{Gyr}$, which is $12.7\text{Gyr}$ assuming $h=0.7$ (for the
$\Lambda$CDM model with the same $\Omega_{m,o}$, the corresponding $t_{o}=13.5\text{Gyr}$). The dimensionless age
parameter $H_{o}t_{o}$ as a function of $\vartheta$ is given in Fig. 1 for various values of $\Omega_{m,o}$
in order to show the sensitivity on this parameter. The inclusion of the radiation
component or the mirage radiation is expected to affect the age very slightly. Of course, an analysis of the model
based on the fittings to the real data (BBN, Supernovae, CMB, BAO) would provide the optimum values of
$\Omega_{m,o}$, $\tilde{V}$ (i.e. $\vartheta$) and therefore the values of $t_{o},z_{p}$. We note that the
normalization to the present value of the Hubble parameter $H_{o}$ has left only one essential free parameter,
$\vartheta$, or to say it differently the parameter $M$ has been traded for $H_{o}$ (of course, there is also
$\Omega_{m,o}$, but this is basically the integration constant of the conservation equation).

The next thing to be found for the matter era is the equation of state for the dark energy. From equations
(\ref{ram}), (\ref{H}) we can find the following equation in terms of the variable $\textrm{y}$
\begin{equation}
2\dot{H}+3H^{2}=-(6M^{3}\gamma)^{2}\Big[\frac{1}{\sqrt{\varphi}}(\sqrt{\varphi}+2\eta\textrm{y}-4)
(\sqrt{\varphi}+\eta\textrm{y}+1)^{2}+\frac{108}{\tilde{V}^{2}}\Big]-\frac{k}{a^{2}}-\frac{\mathcal{C}}{a^{4}}\,.
\label{pouli}
\end{equation}
Comparing equation (\ref{pouli}) with the usual FRW equation $2\dot{H}+3H^{2}=-8\pi G_{N}(p+p_{DE})-\frac
{k}{a^{2}}$, we can define an effective dark energy pressure $p_{DE}$ ($\mathcal{C}/a^{4}$ is ignored)
\begin{equation}
\frac{1}{6M^{6}\gamma}p_{DE}=\frac{1}{\sqrt{\varphi}}(\sqrt{\varphi}+2\eta\textrm{y}-4)
(\sqrt{\varphi}+\eta\textrm{y}+1)^{2}+\frac{108}{\tilde{V}^{2}}\,,
\label{pDE}
\end{equation}
while from (\ref{OmegaDE}) and the usual FRW equation $H^{2}+\frac{k}{a^{2}}=\frac{8\pi G_{N}}{3}
(\rho+\rho_{DE})$, we get the effective dark energy density $\rho_{DE}$
\begin{equation}
\frac{1}{18M^{6}\gamma}\rho_{DE}=\big(\sqrt{\varphi}+\eta\textrm{y}+1\big)^{2}-8\eta\textrm{y}
-\frac{36}{\tilde{V}^{2}}\,.
\label{rhoDE}
\end{equation}
Therefore, in the scenario at hand the dark energy equation of state $w_{DE}=\frac{p_{DE}}{\rho_{DE}}$ is given by
\begin{equation}
w_{DE}=\frac{1}{3\sqrt{\varphi}}
\frac{(\sqrt{\varphi}+2\eta\textrm{y}-4)
(\sqrt{\varphi}+\eta\textrm{y}+1)^{2}+\frac{108}{\tilde{V}^{2}}\sqrt{\varphi}}
{\big(\sqrt{\varphi}+\eta\textrm{y}+1\big)^{2}-8\eta\textrm{y}-\frac{36}{\tilde{V}^{2}}}\,.
\label{eos}
\end{equation}
For the fine-tuned case $\tilde{V}=3+v$ and $z<<10^{7}[M(\text{TeV})]^{3/2}$, we obtain the expression
\begin{equation}
w_{DE}(z)\approx \frac{1}{3}\frac{[(1\!-\!\vartheta)\omega_{_{\!\!\!\!\!\thicksim}o}\!-\!2]
(1+z)^{4}-3\vartheta\omega_{_{\!\!\!\!\!\thicksim}o}}
{[(1\!-\!\vartheta)\omega_{_{\!\!\!\!\!\thicksim}o}\!-\!2]
(1+z)^{4}+\vartheta\omega_{_{\!\!\!\!\!\thicksim}o}}\,.
\label{weff}
\end{equation}
Then, the today value is $w_{DE,o}\approx\frac{1}{3}-\frac{4}{3}\vartheta\big(1-\frac{2}{\omega_{_{\!\!\!\!\!\thicksim}o}}
\big)^{-1}>-1$. For $\vartheta$ tending to the maximum value $1-\frac{2}{\omega_{_{\!\!\!\!\!\thicksim}o}}$, it
is $w_{DE,o}\rightarrow -1$. E.g. for $\vartheta=0.8$, $\Omega_{m,o}=0.3$, it is $w_{DE,o}\approx-0.96$. The function
$w_{DE}(z)$ for $\vartheta=0.8$ is shown in Fig. 2 for various values of $\Omega_{m,o}$. In the far future where
$z\rightarrow -1$, the function $w_{DE}(z)$ always approaches the value $-1$ which corresponds to the cosmological
constant, as expected from equation (\ref{linear1}). Note that the effective dark energy density and pressure are
written as $\rho_{DE}\approx 36M^{6}\gamma\eta\big{\{}[(1\!-\!\vartheta)\omega_{_{\!\!\!\!\!\thicksim}o}-2]\textrm{y}^{4/3}\!+\!\vartheta
\omega_{_{\!\!\!\!\!\thicksim}o}\big{\}}$, $p_{DE}\approx 12M^{6}\gamma\eta
\big{\{}[(1\!-\!\vartheta)\omega_{_{\!\!\!\!\!\thicksim}o}-2]\textrm{y}^{4/3}\!-\!3\vartheta
\omega_{_{\!\!\!\!\!\thicksim}o}\big{\}}$,
and therefore, these components satisfy the weak and dominant energy conditions, while the strong energy condition
is violated for
$\textrm{y}^{4/3}<\vartheta\omega_{_{\!\!\!\!\!\thicksim}o}/[(1\!-\!\vartheta)\omega_{_{\!\!\!\!\!\thicksim}o}-2]$.

\begin{figure}[h!]
\begin{tabular}{cc}
\hspace{-0.5cm}\includegraphics*[width=200pt, height=150pt]{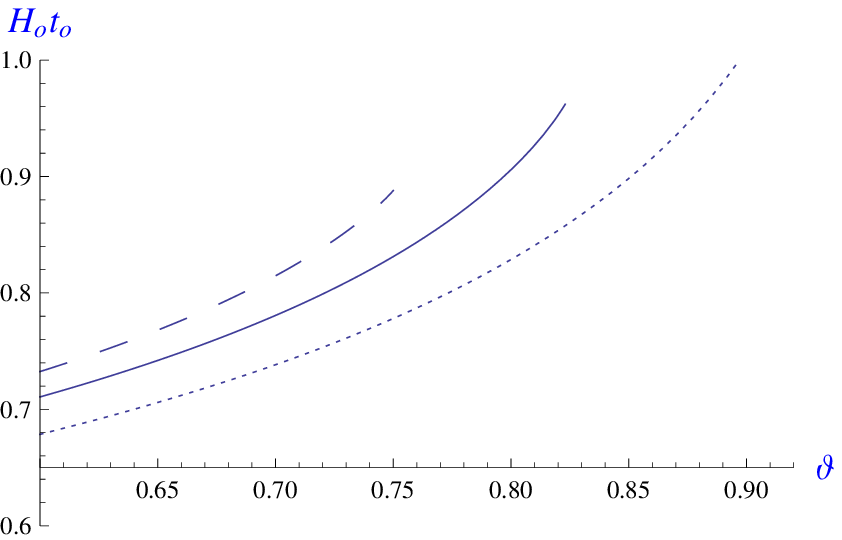}&
\hspace{0.4cm}
\includegraphics*[width=200pt, height=150pt]{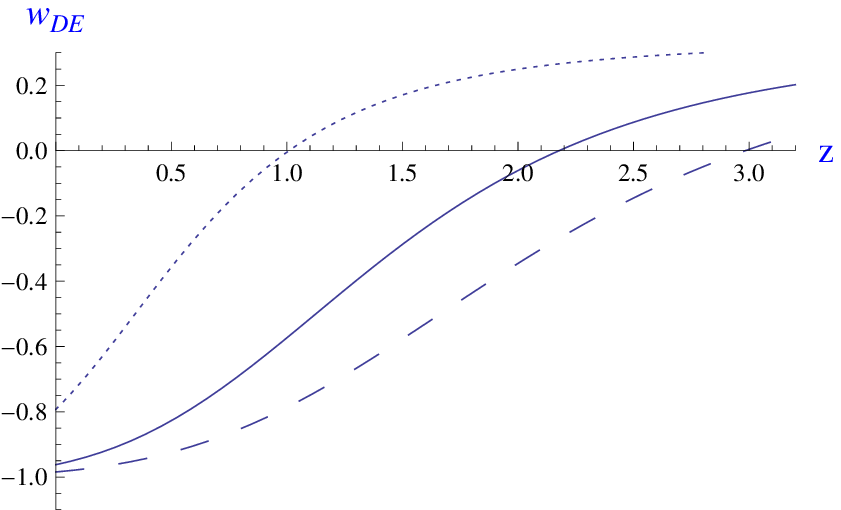}\\
\begin{tabular}{ll}
\hspace{-0.2cm}FIG. 1 : Dimensionless age parameter $H_{o}t_{o}$ as a\\
\hspace{-0.2cm}function of $\vartheta$ for some selected values of $\Omega_{m,o}$. The\\
\hspace{-0.2cm}dotted line corresponds to $\Omega_{m,o}=0.1$, the continuous\\
\hspace{-0.2cm}to 0.3 and the dashed to 0.4
\end{tabular} &
\begin{tabular}{ll}
\hspace{0.8cm}FIG. 2: Dark energy equation of state $w_{DE}$ as a\\
\hspace{0.8cm}function of z for $\vartheta=0.8$ for some selected values of\\
\hspace{0.8cm}$\Omega_{m,o}$. The dotted line corresponds to $\Omega_{m,o}=0.1$,\\
\hspace{0.8cm}the continuous to 0.3 and the dashed to 0.32
\end{tabular}
\end{tabular}
\end{figure}

Let us discuss in brief the situation of the standard matching conditions. Equations (\ref{gala}), (\ref{xara})
are replaced by $\frac{1}{(6M^{3}\gamma)^{2}}\big(H^{2}+\frac{k}{a^{2}}-
\frac{\mathcal{C}}{a^{4}}\big)=4(1+\eta\,\textrm{y})^{2}-\frac{36}{\tilde{V}^{2}}$,
$\frac{1}{(6M^{3}\gamma)^{2}}\frac{\ddot{a}}{a}=4(1-\frac{9}{\tilde{V}^{2}})\!-\!4\eta\textrm{y}
(1+2\eta\textrm{y})$ which still contain the two parameters $\tilde{V},M$ and $\Omega_{m,o}$. However, to
incorporate the flatness values $\Omega_{m,o},\Omega_{DE,o}$ in these standard equations, the relation
$\frac{\Omega_{DE,o}}{\Omega_{m,o}}=\frac{1}{2\eta}\big(1-\frac{9}{\tilde{V}^{2}}+\eta^{2}\big)$ must be
satisfied, which means that one of the two parameters is constrained by $\tilde{V}\approx 3+3\frac{\Omega_{DE,o}}
{\Omega_{m,o}}\eta$, and finally only one parameter remains free. This value of $\tilde{V}$ means that the
standard matching conditions correspond to the limiting value
$\vartheta=1-\frac{2}{\omega_{_{\!\!\!\!\!\thicksim}o}}$.
If $\eta\textrm{y}<<1$, we find for the standard equations
$\frac{1}{(6M^{3}\gamma)^{2}}\big(H^{2}\!+\!\frac{k}{a^{2}}\!-\!
\frac{\mathcal{C}}{a^{4}}\big)\approx 8\big(\frac{\Omega_{DE,o}}{\Omega_{m,o}}+\textrm{y}\big)\eta$,
$\frac{1}{(6M^{3}\gamma)^{2}}\frac{\ddot{a}}{a}\approx 4\big(2\frac{\Omega_{DE,o}}{\Omega_{m,o}}-\textrm{y}\big)\eta$.
These mean that the characteristic quadratic energy density term is insignificant at least for a recent era and
this cosmology coincides with the $\Lambda$CDM one in a recent era. Indeed, today it provides acceleration with a
passage from deceleration to acceleration at $z_{p}\approx 0.67$ for $\Omega_{m,o}=0.3$, while the age of the
universe (for $k=\mathcal{C}=0$, $\Omega_{m,o}=0.3$) is found to be
$t_{o}\approx\frac{1}{3H_{o}\sqrt{\Omega_{m,o}}}\int_{1}^{\infty}\frac{d\textrm{y}}
{\textrm{y}}(\frac{\Omega_{DE,o}}{\Omega_{m,o}}+\textrm{y})^{-1/2}\approx 13.5$Gyr.
From equation (\ref{eos}) setting $\sqrt{\varphi}=1+\eta\textrm{y}$, or from equation (\ref{weff}) setting the
above value of $\vartheta$, it is found that $w_{DE}=-1$, i.e. the exact
cosmological constant. In this standard picture there is no essential free parameter left due to the constraint
from the observed Hubble constant $H_{o}$ ($\Omega_{m,o}$ is still as always free, but approximately known).
As a result, the proposed cosmology, compared to the standard one, possesses more freedom in accommodating the
observed characteristics of the universe, e.g. age of the universe, recent passage from deceleration to
acceleration, time variability of the dark energy equation of state, etc.

{\underline{\textit{Radiation era}}} ($w=1/3$): Equation (\ref{aca}) becomes
\begin{equation}
\frac{1}{(6M^{3}\gamma)^{2}}\frac{\ddot{a}}{a}=\frac{1}{\sqrt{\varphi}}\big(2-\sqrt{\varphi}-
2\eta\,\textrm{y}\big)\big(1+\sqrt{\varphi}+\eta\,\textrm{y}\big)^{\!2}-\frac{36}{\tilde{V}^{2}}-
\frac{1}{(6M^{3}\gamma)^{2}}\,\frac{\mathcal{C}}{a^{4}}\,,
\label{radixara}
\end{equation}
where the constant $\mathcal{C}$ now cannot be ignored at early times, and
\begin{equation}
\varphi=(1+\eta\,\textrm{y})^{2}+\frac{cx_{o}^{4/3}}{\eta^{1/3}\tilde{V}^{7/3}}\textrm{y}\,.
\label{raditan}
\end{equation}
We will discuss the fine-tuned case $\tilde{V}=3+v$, where it is important to note that the same parameters
which gave the late-times evolution will be used also for the early-times behaviour. So, the
value of $cx_{o}^{4/3}$ found at late times will be used also here. This way, we will have a unified,
all-times cosmology. For $\eta\textrm{y}^{3/2}>>1$, i.e. $z>>10^{5}M(\text{TeV})$, it is
$\varphi\approx(\eta\textrm{y})^{2}
+\frac{1}{\sqrt[3]{3}}[(1-\vartheta)\omega_{_{\!\!\!\!\!\thicksim}o}-2]\eta^{2/3}\textrm{y}$. More precisely,
for $10^{5}M(\text{TeV})\!<<\!z\!<<\!10^{10}[M(\text{TeV})]^{2}$ it is $\varphi\approx
\frac{1}{\sqrt[3]{3}}[(1-\vartheta)\omega_{_{\!\!\!\!\!\thicksim}o}-2]\eta^{2/3}\textrm{y}$ and
$\frac{\ddot{a}}{a}\approx-(6M^{3}\gamma)^{2}\varphi-\frac{\mathcal{C}}{a^{4}}
\approx-\frac{1}{\sqrt[3]{3}}(6M^{3}\gamma)^{2}
[(1-\vartheta)\omega_{_{\!\!\!\!\!\thicksim}o}-2]\eta^{2/3}z^{4}-\mathcal{C}z^{4}$;
for $z\sim 10^{10}[M(\text{TeV})]^{2}$ both terms are significant in (\ref{raditan}) and
$\frac{\ddot{a}}{a}\approx-\frac{(6M^{3}\gamma)^{2}}{\sqrt{\varphi}}(\sqrt{\varphi}+2\eta\textrm{y})
(\sqrt{\varphi}+\eta\textrm{y})^{2}-\frac{\mathcal{C}}{a^{4}}$; for
$z\!>>\!10^{10}[M(\text{TeV})]^{2}$ it is $\varphi\approx(\eta\textrm{y})^{2}$ and
$\frac{\ddot{a}}{a}\approx-12(6M^{3}\gamma)^{2}\varphi-\frac{\mathcal{C}}{a^{4}}
\approx-12(6M^{3}\gamma)^{2}\eta^{2}z^{8}-\mathcal{C}z^{4}$. This last case is the most interesting one since
it can encapsulate an inflationary accelerating period,
where for temperature of inflation $T_{inf}\sim 1\text{TeV}$ it is
$z_{inf}\approx\frac{T_{inf}}{T_{o}}\sim 10^{16}$ (these refer to the end of inflation),
while for $T_{inf}\sim M_{Pl}$ it is $z_{inf}\sim 10^{29}$. It is obvious that initially the dominant power
in the acceleration expression is $z^{8}$, which has negative sign, therefore the universe starts with
deceleration. However, if $\mathcal{C}<0$, the universe necessarily enters an accelerating phase during the
evolution. To make a more precise estimate, if $z_{ent}$ denotes the entrance into the accelerating phase, it is
$|\mathcal{C}|\approx 12(6M^{3}\gamma)^{2}\eta^{2}z_{ent}^{4}\approx 10^{-120}[M(\text{TeV})]^{-6}
z_{ent}^{4}\text{TeV}^{2}$. If the accelerated phase is to be interpreted as the inflationary era, the redshift
$z_{ent}$ should be several orders of magnitude larger than $z_{inf}$. Since the maximum acceleration occurs
at redshift $z_{max}=2^{-1/4}z_{ent}$, the value of this maximum acceleration is
$\frac{\ddot{a}}{a}|_{max}=3(6M^{3}\gamma)^{2}\eta^{2}z_{ent}^{8}$ and its ratio to the today acceleration
is $\frac{\ddot{a}}{a}|_{max}/\frac{\ddot{a}}{a}|_{o}\sim\eta z_{ent}^{8}$. This ratio is various decades of
order of magnitude bigger than 1, which means that the early acceleration is huge compared to the today tiny
value. Although this is interesting, there is a caveat. Since $H_{o}^{2}\approx 10^{-90}h^{2}\text{TeV}^{2}$, for
$z_{ent}>>10^{7.5}[M(\text{TeV})]^{1.5}$ it arises $|\mathcal{C}|>>H_{o}^{2}$.
On the other hand, from equation (\ref{flat}) it is $H^{2}+\frac{k}{a^{2}}=\frac{8\pi G_{N}}{3}
(\rho+\rho_{DE}+\rho_{\mathcal{C}})$, where the energy density of dark radiation is
$\rho_{\mathcal{C}}=\frac{3\mathcal{C}}{8\pi G_{N} a^{4}}$. In order not to violate the nucleosynthesis bounds it
should be $-1.23\leq\frac{\rho_{\mathcal{C}}}{\rho_{r}}\leq 0.11$ \cite{nucl}, where $\rho_{r}$ is the energy
density of the radiation component. Since $\rho_{r}=\frac{3\Omega_{r,o} H_{o}^{2}}{8\pi G_{N}a^{4}}$,
we get $-1.23\leq\frac{\mathcal{C}}{\Omega_{r,o}H_{o}^{2}}\leq 0.11$, which means that $|\mathcal{C}|<H_{o}^{2}$.
Therefore, there is a contradiction. Although a sufficient negative value of $\mathcal{C}$ leads to a large
early-times acceleration, in order for this phase to be interpreted as inflation the value of $\mathcal{C}$
should violate the nucleosynthesis constraints.

Other values of $\tilde{V}$, different than $\tilde{V}=3+v$,
may also be interesting for providing a geometric origin inflationary period in the early cosmic era, but
this investigation is beyond the scope of the present study. For example, assuming a varying tension in the early
universe, it might be possible to unify a large value of $\tilde{V}$ in the inflationary period with the above
fine-tuned value of $\tilde{V}$ which provides a solution to the dark energy problem. The physical motivation
behind this dependence of the brane tension can be ideas inspired by the temperature dependence of the fluid
membrane tension, cosmological phase transitions that modify brane tension, brane-bulk energy exchange, or particle
creation on the brane \cite{abdalla}. A hybrid type of inflation could be used with the varying tension being the
``field" that produces the exponential expansion while an extra scalar field could cause the end of inflation and
reheating.

\section{Conclusions} \label{Conclusions}

In this paper, we continue the investigation of a recent proposal \cite{kof-ira} on alternative matching
conditions for self-gravitating branes. The bulk metric is assumed to be regular at the brane position, as
e.g. happens in the braneworld scenario. While an equation of the general form bulk gravity tensor equals some
smooth matter content or some matter content of a ``thick'' brane is certainly correct, we claim that it cannot
be correct in the shrink limit of distributional branes. A different treatment of the delta function
characterizing the defect is needed for extracting its equation of motion. If this is so, the Israel matching
conditions, as well as their generalizations where the bulk gravity tensor instead of Einstein is replaced by
Lovelock extensions and the branes have differing codimensions, may be physically inadequate.

Our reasoning is based on two points: First, the incapability of the conventional matching conditions to
accept the Nambu-Goto probe limit. Even the geodesic limit of the Israel matching conditions is not an
acceptable probe limit since being the geodesic equation a kinematical fact it should be preserved independent
of the gravitational theory or the codimension of the defect, which however is not the case for these matching
conditions. Furthermore, even the non-geodesic probe limit of the standard equations of motion for various codimension
defects in Lovelock gravity theories is not accepted, since this consists of higher order algebraic equations in the
extrinsic curvature, therefore, a multiplicity of probe solutions arise instead of a unique equation of motion at
the probe level. Second, in the $D$-dimensional spacetime we live (maybe $D=4$), classical defects of any possible
codimension could in principle be constructed, and therefore, they should be compatible. The
standard matching conditions fail to accept codimension-2 and 3 defects for $D=4$ (which represents effectively
the spacetime at certain length and energy scales) and most probably fail to accept high enough codimensional
defects for any $D$ since there is no corresponding high enough Lovelock density to support them.

According to our proposal the problem is not the distributional character of the defects, neither the
gravitational theory used, but the equations of motion of the defects. The proposed matching conditions
(``gravitating Nambu-Goto matching conditions'') may be close to the correct direction of finding realistic
matching conditions since they always have the Nambu-Goto probe limit (independently of the gravity theory, the
dimensionality of spacetime or codimensionality of the brane), and moreover, with these
matching conditions, defects of any codimension seem to be consistent for any (second order) gravity theory.
These alternative matching conditions arise by promoting the embedding fields of the defect to the
fundamental entities. Instead of varying the brane-bulk action with respect to the bulk metric at the brane
position and derive the standard matching conditions, we vary with respect to the brane embedding fields
in a way that takes into account the gravitational back-reaction of the brane to the bulk. The proposed matching
conditions generalize the standard matching conditions, and so, all the solutions of the bulk equations of
motion plus the conventional matching conditions are still solutions of the current system of equations.
Therefore, only interesting extensions are expected by using the proposed matching conditions.

In the present work we have considered in detail the case of a 3-brane in five-dimensional Einstein gravity and
derived the generic alternative matching conditions. Of course, same or similar results are true for other
codimension-1 defects in other spacetime dimensions. Since a 3-brane can represent our world in the braneworld
scenario, we have investigated the cosmological equations and found the general solution for the cosmic evolution,
as well as its bulk extension. One branch of the solution, that we have investigated further, possesses the
asymptotic linearized LFRW regime. Compared to the conventional 5-dimensional braneworld cosmology, here, the
Friedmann equation is much more complicated and has an extra integration constant.

Both in the standard and the alternative cosmologies, there are three parameters in the action: the higher
dimensional mass scale, the bulk cosmological constant and the brane tension. Since both cosmologies possess the
linearized asymptotic regime, the satisfaction of Newton's constant constrains the parameters from three
to two. The main difference between the two cosmologies is the satisfaction of the (approximately known)
today flatness parameters. For the standard braneworld cosmology (which coincides with $\Lambda$CDM at least recently)
this requirement is achieved in charge of one of these two parameters, and finally, only one parameter remains
free (actually no essential free parameter remains due to the normalization to the present Hubble constant).
On the contrary, in the proposed cosmology, the existence of the extra integration constant accounts for the
today matter and dark energy contents, and finally, the two parameters remain free (again only one free parameter
is essential due to the today Hubble value, and this parameter is denoted in the paper by $\tilde{V}$ or
$\vartheta$). Therefore, the new cosmology possesses an extra freedom for accommodating better the observed
characteristics of the universe. We have found that for values of $\tilde{V}$ extremely close to the
Randall-Sundrum fine-tuning there is a small today acceleration with a recent passage from the long deceleration
era to the present epoch. We have estimated the age of the universe which is consistent with current data, and
calculated the time variability of the dark energy equation of state. For the same values of $\tilde{V}$ a
unified cosmology is defined for all times which possesses in the radiation regime a large acceleration (however,
this cannot easily be interpreted as inflation since the nucleosynthesis bounds are violated). In general,
depending on the parameters, a variety of behaviours can be exhibited which need further investigation and the
model should be confronted against real data.

\[ \]
{{\bf Acknowlegements}}
We wish to thank E. Saridakis and J. Zanelli for useful discussions.

\appendix

\section{Geometric Components} \label{geometric components}

For the metric (\ref{metric}), the extrinsic curvature tensor defined everywhere in the bulk is
$\mathcal{K}_{ij}(\chi,y)=\frac{1}{2}g_{ij}^{\prime}(\chi,y)$ (a prime denotes $\partial/\partial y$).
In the text we denote the two side value of $\mathcal{K}_{ij}$ at the brane position by
$K_{ij}^{+}(\chi)\equiv \mathcal{K}_{ij}(\chi,0^{+})$,
$K_{ij}^{-}(\chi)\equiv \mathcal{K}_{ij}(\chi,0^{-})$.
The non-vanishing components of the necessary geometric quantities are
\begin{equation}
\Gamma^{5}_{ij}\!=\!-\mathcal{K}_{ij}\,\,\,\,\,,\,\,\,\,\,
\Gamma^{j}_{i5}\!=\!\mathcal{K}^{j}_{i}\,\,\,\,\,,\,\,\,\,\,
\Gamma^{i}_{jk}\!=\!\frac{1}{2}g^{i\ell}(g_{\ell j,k}\!+\!g_{\ell k,j}\!-\!g_{jk,\ell})
\label{non-vanishing Christoffels}
\end{equation}
(where $\mathcal{K}^{i}_{j}=g^{ik}\mathcal{K}_{kj}$)
\begin{equation}
\mathcal{R}_{i5j5}=-\mathcal{K}_{ij}^{\prime}+\mathcal{K}_{ik}\mathcal{K}_{j}^{k}\,\,\,\,\,,\,\,\,\,\,
\mathcal{R}_{5ijk}=\mathcal{K}_{ij|k}-\mathcal{K}_{ik|j}\,\,\,\,\,,\,\,\,\,\,
\mathcal{R}_{ijk\ell}=R_{ijk\ell}+\mathcal{K}_{i\ell}\mathcal{K}_{jk}-\mathcal{K}_{ik}\mathcal{K}_{j\ell}
\label{non-vanishing Riemanns}
\end{equation}
(where $|$ denotes the covariant derivative with respect to the metric $g_{ij}$)
\begin{equation}
\mathcal{R}_{55}=-\mathcal{K}^{\prime}-\mathcal{K}_{ij}\mathcal{K}^{ij}\,\,\,\,\,,\,\,\,\,\,
\mathcal{R}_{i5}=\mathcal{K}^{j}_{i|j}-\mathcal{K}_{|i}\,\,\,\,\,,\,\,\,\,\,
\mathcal{R}_{ij}=R_{ij}-\mathcal{K}'_{ij}+2\mathcal{K}_{ik}\mathcal{K}_{j}^{k}-\mathcal{K}\mathcal{K}_{ij}
\label{non-vanishing Riccis}
\end{equation}
(where $\mathcal{K}=\mathcal{K}_{i}^{i}$)
\begin{equation}
\mathcal{R}=R-2\mathcal{K}'-\mathcal{K}_{ij}\mathcal{K}^{ij}-\mathcal{K}^{2}
\label{Ricci scalar}
\end{equation}
\begin{equation}
\mathcal{G}_{55}=\frac{1}{2} \mathcal{K}^{2}-\frac{1}{2}\mathcal{K}_{ij}\mathcal{K}^{ij}-\frac{1}{2}R
\,\,\,\,,\,\,\,\,
\mathcal{G}_{i5}=\mathcal{K}^{j}_{i|j}-\mathcal{K}_{|i}\,\,\,\,,\,\,\,\,
\mathcal{G}_{ij}=G_{ij}-\mathcal{K}'_{ij}+2\mathcal{K}_{ik}\mathcal{K}_{j}^{k}-\mathcal{K}\mathcal{K}_{ij}
+\Big(\mathcal{K}'+\frac{1}{2}\mathcal{K}_{k\ell}\mathcal{K}^{k\ell}+\frac{1}{2}\mathcal{K}^{2}\Big)g_{ij}\,.
\label{non-vanishing Einsteins}
\end{equation}


\end{document}